\documentclass[useAMS,usenatbib]{mn2e}
\input{psfig}
\usepackage{graphicx}
\usepackage{rotating}
\usepackage{natbib}
\usepackage{amssymb}
\usepackage{amsmath}
\usepackage{aas_macros}
\usepackage{fixltx2e}

\title
[Central image in gravitational lens]
{A search for the third lensed image in JVAS B1030+074}

\author
[M.~Zhang et al.]
{
M.~Zhang,$^1$
\thanks{E-mail: mzhang@jb.man.ac.uk}
N.~Jackson,$^1$  R.~W.~Porcas$^2$ and I.~W.~A.~Browne$^1$ \\
$^1$University of Manchester, Jodrell Bank Observatory, Macclesfield,
Cheshire SK11 9DL\\ $^2$Max-Planck-Institut f\"{u}r Radioastronomie,
Auf dem H\"{u}gel 69, D 53121, Bonn, Germany\\
}

\def\ltsim  {\ifmmode\stackrel{<}{_{\sim}}\else$\stackrel{<}{_{\sim}}$\fi}
\def\gtsim  {\ifmmode\stackrel{>}{_{\sim}}\else$\stackrel{>}{_{\sim}}$\fi}

\def\mujybm {${\rm \mu}$Jy\,beam$^{-1}$}
\begin{document}

\maketitle

\begin{abstract} 
{Central gravitational image detection is very important for the study
  of the mass distribution of the inner parts ($\sim 100$~pc) of lens
  galaxies. However, the detection of such images is extremely rare
  and difficult. We present a 1.7-GHz High Sensitivity Array (HSA)
  observation of the double-image radio lens system B1030+074. The
  data are combined with archive VLBA and global-VLBI observations,
  and careful consideration is given to the effects of noise, {\sc
  clean}ing and self-calibration. An upper limit is derived for the
  strength of the central image of 180~$\mu$Jy (90\% confidence
  level), considerably greater than would have been expected on the
  basis of a simple analysis.  This gives a lower limit of $\sim 10^3$
  for the ratio of the brightest image to the central image.  For
  cusped models of lens mass distributions, we have made use of this
  non-detection to constrain the relation between inner power-law
  slope $\beta$ of the lensing galaxy mass profile, and its break
  radius $r_b$. For $r_b>130$~pc the power-law slope is required to be
  close to isothermal ($\beta>1.8$).  A flatter inner slope is allowed
  if a massive black hole is present at the centre of the lensing
  galaxy, but the effect of the black hole is small unless it is
  $\sim 10$ times more massive than that implied by the relation
  between black hole mass and stellar velocity dispersion.  By
  comparing four epochs of VLBI observations, we also detected
  possible superluminal motion in the jet in the brighter A image.
  The B jet remains unresolved, as expected from a simple lens model
  of the system.}
\end{abstract}


\section{Introduction}

Gravitational lensing is an important method to probe the mass
distribution in the Universe, from the largest scales down to
sub-galactic scales. The matter in our universe is thought to be
predominantly non-baryonic, consisting of a weakly-interacting form
known as Cold Dark Matter (CDM). CDM simulations have so far been very
successful in reproducing large-scale structures observed in large
galaxy surveys such as the 2dF \citep{colless.01.mn,peacock.03.aip}
and SDSS \citep{doroshkevich.04.a&a}. CDM structures form potential
wells, and numerical simulations predict that these have a
characteristic radial mass profile consisting of a broken power law
with a flatter central slope, now known as a NFW profile
\citep*{navarro.96.apj,navarro.97.apj}.  Baryons dominate the central
regions of galaxies, having cooled into the centres of the potential
wells formed by the CDM
\citep*{rees.77.mn,navarro.91.apj,combes.04.iau}. However, the detailed
physics is difficult; firstly, large-scale CDM simulations reach the
limit of their current resolution on scales of $\sim$~kpc
\citep[e.g.][]{ghigna.00.apj,sommer-larsen.03.apj,robertson.04.apj,kang.05.apj}
and second, the baryon content requires the input of detailed
micro-physics \citep[e.g.][]{white.91.apj,gnedin.04.apj,maccio.06.mn}.

In the central regions of galaxies we have reasons to believe that an
additional component is present, namely a massive black hole. At the
centre of our own Galaxy there is strong evidence for the existence of
a $3.7\times10^6~\rm M_{\odot}$ black hole \citep{ghez.05.apj}, and
more massive black holes of $\sim 0.002$ times the total galaxy mass
have been inferred from kinematical observations
\citep{lauer.97.asp}. Some galaxies have black hole masses as large as
$1\times10^9~\rm M_\odot$ \citep{kormendy.96.apj}. Radio observations
have also shown that massive black holes exist in active disk galaxies
\citep{miyoshi.95.nat}. The black hole mass is also found to scale
with the stellar velocity dispersion of the bulge
\citep{ferrarese.00.apj,gebhardt.00.apj,tremaine.02.apj}, in a process
which is probably related to galaxy formation in a way that is not yet
completely understood.

The mass distribution in the central regions of galaxies can be probed
by gravitational lensing, as was first pointed out by
\citet*{wallington.93.apj}. This is because the central image of a
gravitational lens system passes through the centre of the lens galaxy
potential, and its magnification can in principle tell us about the
nature of the potential within $\sim 100$~pc from the centre of the
galaxy. In general, the more nearly singular the potential, the more
strongly demagnified is the central image.

A number of theoretical and observational analyses of central images
of galaxy lens systems have been carried out to date.  A strict proof
for the general theorem that lenses have odd numbers of images was
carried out with Lorentzian manifolds \citep{mckenzie.85.jmp}, though
its applicable conditions are arguably not universal
\citep{gottlieb.94.jmp}. However, most physical models show the
existence of the central odd image (Section
\ref{model}). Observationally, there are some cases in which radio
emission is detected near the centre of a lens galaxy, for example in
the radio-loud lens systems MG~J1131+0456 \citep*{chen.93.aj}, QSO
0957+561 \citep{harvanek.97.aj}, CLASS B2045+265
\citep{fassnacht.99.aj} and CLASS B2108+213 \citep{mckean.05.mn}. None
of these are secure or even probable detections of a central lensed
image, as the central component is quite likely to be due to radio
emission from active nuclei in the lens galaxies. The lens system PMN
J1632$-$0033 has yielded the most likely detection of a central image,
although in this case free-free absorption in the lensing galaxy is
required to modify the radio spectral index of the central image to
make it agree with that of the other images
\citep{winn.03.apj,winn.04.nat}. In addition, there is one
optically-selected three-image case, APM~08279+5255
\citep{ibata.99.aj,egami.00.apj,munoz.01.apj}, which may also be
caused by a ``naked cusp'' configuration rather than a central minimum
\citep{lewis.02.mn}. A likely central fifth image has recently been
discovered in SDSS J1004+4112 in which the lens is a cluster of
galaxies \citep{inada.05.pasj}. Recently the more systematic
Extragalactic Lens VLBI Imaging Survey (ELVIS) project has attempted
to find detections or limits on core images, resulting in one
non-detection so far in PMN J1838-3427 \citep{boyce.06.apj}.
 
There are different views on the mass distribution in the inner
regions of galaxies. Existing observations can be used to constrain
the mass distribution of central regions of the lensing galaxy. Since
numerical simulations predict double power-law models with central
cusps rather than core radii \citep{navarro.97.apj}, and the observed
light distributions seem also to follow cusp-like laws in the centres
of nearby early-type galaxies \citep{faber.97.aj}, these analytic
forms are usually assumed when fitting central mass distributions
\citep{munoz.01.apj,rusin.03.apj} On the other hand, in some cases,
in particular in low-surface-brightness galaxies where a large dark
matter component may be present, \citet{de_blok.01.apj} have argued
that the central power-law exponent is close to flat, corresponding to
a central density ``core''. Central image detection (or limits) can
help us to understand the mass distribution.

A thorough analysis of the available constraints from the non-detection
of the central image has been carried out by \citet{keeton.03.apj} and
\citet{mao.01.mn} have extended the analysis to cover the properties of
central black holes. In principle, a non-detection of a central image
implies either a relatively steep inner mass slope or a small break
radius in the power-law. \citet{rusin.01.apj}, based on analysis of six
systems, found that the inner mass slope index is unlikely to be flatter
than 0.8. \citet{keeton.03.apj} discussed a range of models for
realistic lens galaxies and found that galaxies which yield central
image magnifications of $\mu\sim 0.001$ are likely to be the most
common, and that in 10--20\% of cases a central black hole may be large
enough, and positioned close enough to the central image, to demagnify
it into invisibility.

In this paper we consider the lens system B1030+074
\citep{xanthopoulos.98.mn}. This is in many ways the ideal system in
which to look for an odd image. First, it is very strong, having a
primary component with flux density of more than 200~mJy at 5~GHz
\citep{xanthopoulos.98.mn}. Second, being a radio source, the
detection of the central image is made easier by the absence of strong
radio emission from the lensing galaxy.  Third, it has two observed
images which are relatively asymmetric in flux density
($\sim 15:1$). Such lens systems are likely to give rise to central
images which are less strongly demagnified than four-image lens
systems or more symmetric two-image systems \citep{winn.03.apj}.  Our
aim was to produce an image with an rms noise of about $10^{-4}$ of
the flux of the primary, and hence give us a reasonable chance to
detect the central image based on the prediction of
\citet{keeton.03.apj}. In the following Section \ref{observ},
\ref{noise}, \ref{model} and \ref{superlum}, we present results from
observations, a derivation of the upper limit to detection, results
from lens modelling and detection of superluminal jet motions,
respectively.

\section{Observations}\label{observ}

The double-image lens system B1030+074 was discovered during the
8.4-GHz Jodrell-VLA Astrometric Survey (JVAS) in 1992
\citep{browne.98.mn,xanthopoulos.98.mn}.  A follow-up 1.7-GHz
observation with Very Long Baseline Array (VLBA) and Effelsberg showed
the primary image had a flux density of 250~mJy and the secondary
image had a flux density of 17.4~mJy with a separation of 1.567~arcsec
at a position angle of $143^\circ.4$. The nominal dynamic range of the
1.7-GHz map was about 6000:1 \citep{xanthopoulos.01.asp}.  We have
made a new L-band Very Long Baseline Interferometry (VLBI) observation
with the High Sensitivity Array (HSA: VLBA, VLA, Green Bank Telescope
and Arecibo) aimed at approaching a higher dynamic range in order to
detect any central image or improve the upper limit. This 1.7-GHz dual
polarisation observation
%
%
was carried out on 2004 December 27. About 3 hours integration were
obtained with an aggregate bit rate of 256~Mbs$^{-1}$. In frequency
this corresponds to 4 IFs each containing 32 channels of 0.25~MHz,
using 2~bits/sample. The data were correlated using the VLBA
correlator in Socorro, averaged into 2-s bins. A single field centre
was used, centred on the mid point between image A and B.

The data calibration was mainly carried out with {\sc
aips}\footnote{the Astronomical Image Processing System distributed by
NRAO (National Radio Astronomy Observatory)}.  Bad data were flagged,
including the removal of the upper two spectral channels, all
baselines involving the Mauna Kea VLBA antenna, and a large segment of
data from the Green Bank antenna in the first hour of
observations. The VLBA calibration transfer, written by the correlator
into the data file, was applied, having been corrected for an
incorrect system temperature due to a VLA mode single dish/phased
array recognition problem (Sjouwerman 2005, NRAO communication). The
data were re-inspected and further sporadic episodes of bad data were
flagged. A fringe-fit was then performed using the data on the target
source B1030+074; we have experimented with use of a two-component
model and a point source in the fringe fit and find that little
difference is evident. The data after fringe fitting were checked to
ensure a flat response across the bandpass, and the phase and
amplitude solutions derived during the calibration process were
applied to the data.

The data were averaged in time using the {\sc aips} task {\sc ubavg},
which performs a baseline-dependent time averaging, giving $<1$\% flux
reduction by time-average smearing over the 1600-mas extent of the
source. In frequency, the data were then averaged into 8 channels,
each with a bandwidth of 1 MHz, which gives a flux reduction of $<3\%$
over the required field of view. Images were produced using the {\sc
aips imagr} program, which implements the Clark version of the {\sc
clean} algorithm \citep{hogbom.74.a&as,clark.80.a&a}. The data were
self-calibrated using the initial image as a model, taking {\sc clean}
components up to the first negative. A global amplitude
self-calibration was first applied, using a very long solution
interval to adjust the flux scale of the antennas to each other. After
production of a new {\sc clean}ed image, a phase self-calibration was
then applied using a short solution interval. After correction of
phases in this way, a further round of self-calibration using
amplitudes and phases was performed, with a solution interval of 15
minutes. No further improvement was observed in the maps by use of
further self-calibration. Final images were produced using $uv$ grid
weighting close to uniform and weighting of the data points by the
statistical weight appropriate to the size of the telescopes on the
baseline involved. Natural grid weighting in principle should give
better signal-to-noise by a factor of $\sim 2$, but was found to give
very poor results in practice. This is because calibration errors and
{\sc clean} errors are more important in high-dynamic-range mapping,
we cannot expect the thermal noise limit to apply. In general, for
producing the final image from the self-calibrated data we {\sc
clean}ed the whole image, or a large part of it, using a large number
($\sim 20000$) of {\sc clean} iterations, a procedure which we justify
in detail in Section \ref{noise}. Gaussian model fitting to the final
images was done with the {\sc aips} task {\sc jmfit} in the image
plane.

\begin{figure*}
\begin{tabular}{c}
\psfig{figure=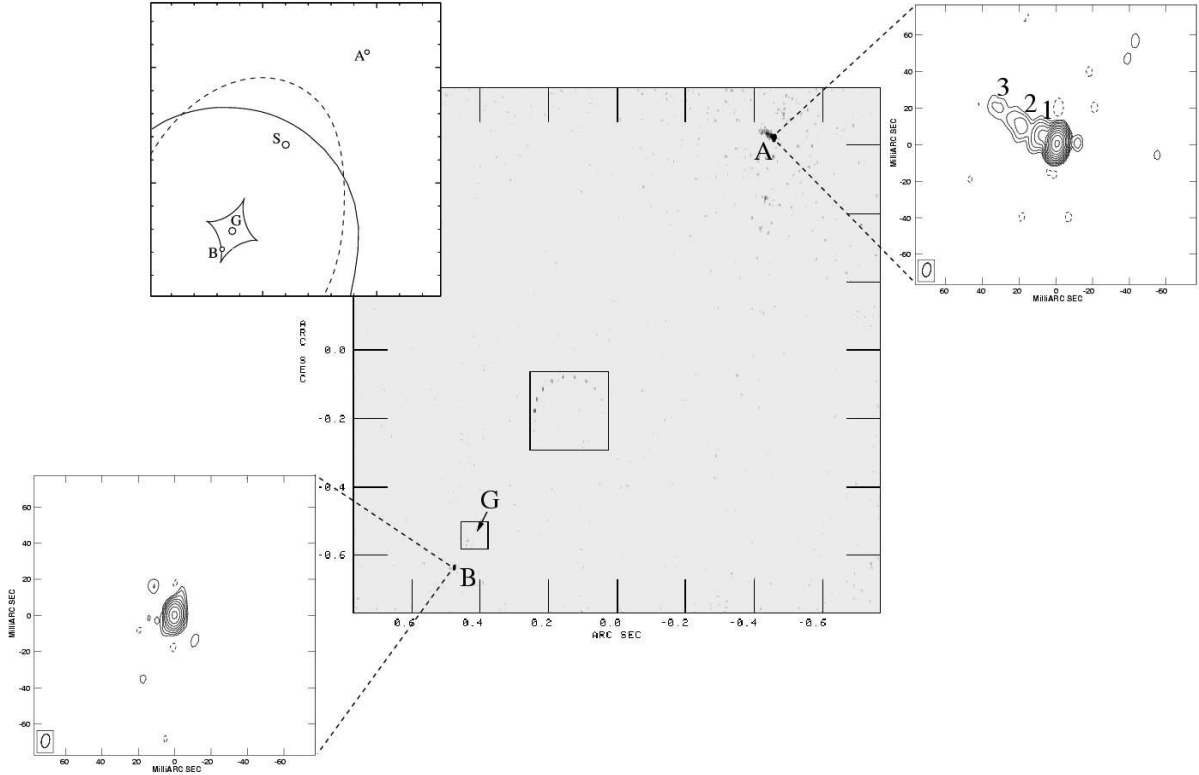,width=16cm} 
\end{tabular}
\caption{Combined global-VLBI + HSA images of B1030+074. The restoring
  beam size is 7.7$\times$4.5~mas$^2$ at a position angle of
  $-10^\circ.9$. Labels A, B, G and S denote primary A, secondary B
  images, lens galaxy and the source respectively.  Image A and image
  B are shown in two separate magnified figures. Contours in the maps
  are plotted at multiples (-1, 1, 2, 4, 8, 16, 32, 64, 128, 256, 512,
  1024, 2048) $\times$ $3\sigma$ where $\sigma$ is the off-source
  local rms noise in the map (53~\mujybm ~in A image and 26~\mujybm
  ~in B image). The top-left figure shows a fitted SIE model, while
  the solid curves are caustics and the dashed one is the critical
  curve. The two boxes in the map are our {\sc clean} recovery test
  region and third image prediction region respectively, see details
  in Section \ref{noise}.}
\label{fig:whole}
\end{figure*}

To get the maximum dynamic range, we also obtained data on B1030+074
from previous 1.7-GHz VLBI observations (1998 BX003, 2000 GX006, 2001
GX007), and combined them with our new observation (2004 BJ054).  Data
from the previous epochs were calibrated and imaged in the same way as
previously described. For each epoch, {\sc clean}ed maps were made
individually, and combined together in the map plane weighting by
$1/\sigma^2$, where $\sigma$ is the measured rms noise in each image.
Our combined global-VLBI + HSA map of B1030+074 made with a restoring
beam of 7.7$\times$4.5~mas$^2$ at a position angle of
$-10^\circ.85$, is shown in Fig.~\ref{fig:whole}, together with
enlarged versions of images A and B. Parameter values from model
fitting are given in Table~\ref{tab:b1030comb}.

Image A is resolved with an extended jet that can be seen in the
north-east direction having a length of about 20 mas, while image B is
not resolved. A super-resolved image with a half-size restoring beam
(i.e., 3.8$\times$4.5~mas$^2$) was made, but the B jet was still not
detected. The whole field {\sc clean} of the HSA map gives a noise
level of $\sigma\sim 24.1$~\mujybm ~around the expected position of the
central image (see Table~\ref{tab:4obnois}). After the combination of
the maps from four epochs, the noise level is reduced to
$\sigma\sim 19.9$~\mujybm. The dynamic range achieved in the final map
is more than 10000:1 ($S_A/\sigma >10000$, see
Table~\ref{tab:b1030comb}).

\begin{table*}
\begin{tabular}{cccccc} \hline
Experiment code & Telescope & Observing date & Time on source &
          Measured rms noise & Thermal noise \\
    &      & (yyyy mm dd) & (min) & (\mujybm) & (\mujybm) \\  \hline
BX003 & VLBA+Eb             & 1998 06 10 & 705.00 &  61.8  & 44.4 \\
GX006 & VLBA+Eb+Jb+Wb       & 2000 02 12 & 545.00 &  58.1  & 24.1 \\
GX007 & VLBA+VLA+EVN subset & 2001 02 09 & 526.00 &  39.5  & 18.9 \\  
BJ054 & HSA                 & 2004 12 27 & 166.25 &  24.1  &  8.2 \\ \hline
\end{tabular}
\caption{Noise levels in the central image regions of four epoch
  1.7-GHz VLBI/HSA observations }
\label{tab:4obnois}
\end{table*}

\begin{table*}
\begin{tabular}{crrrr} \hline
 & R.A. offset & Dec offset & Flux density & Peak intensity \\
 & (mas) & (mas) & (mJy) & (mJy beam$^{-1}$) \\ \hline
A  & 0.0000 & 0.0000  & $258\pm13$ & $208\pm10$\\
A1 & $5.613\pm0.006$ & $3.158\pm0.005$ & $  8.5\pm0.4$ & $  3.6\pm0.2$\\
A2 & $20.47\pm0.03$ & $9.96\pm0.04$ & $  2.6\pm0.1$ & $  0.73\pm0.04$\\
A3 & $30.0\pm0.1$ & $17.4\pm0.1$ & $  1.00\pm0.05$ & $  0.24\pm0.01$\\
B &  $934.987\pm0.001$ & $-1258.092\pm0.002$ & $ 16.9\pm0.8$ & $ 12.5\pm0.6$\\
G &  $869.0\pm10.0$ & $-1147.0\pm10.0$ & - & - \\ \hline
\end{tabular}
\caption{The observational constraints from combined VLBI + HSA
  images. A1, A2 and A3 are jet components within component A
  (top-right figure in Fig.~\ref{fig:whole}) and G is the lensing galaxy. The
  position of G comes from HST I-band image (see text).}
\label{tab:b1030comb}
\end{table*}

\begin{figure}
\psfig{figure=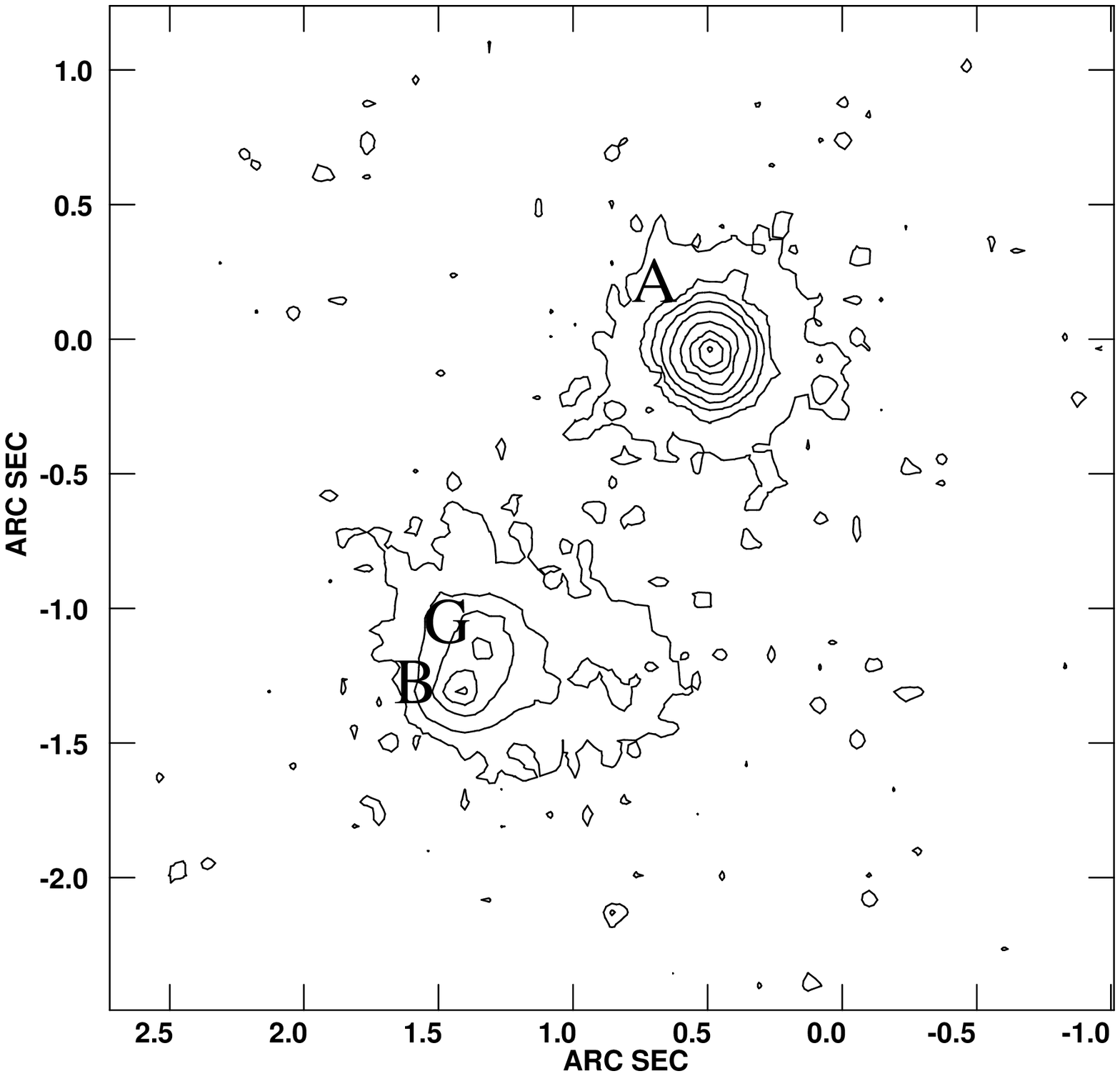,width=8cm}
\caption{HST I-band image of B1030+074. The diffuse double-peak
  pattern at the lower part includes the B image and lens
  galaxy. Contours in the map are plotted at multiples (-1, 1, 2, 4,
  8, 16, 32, 64, 128, 256, 512) $\times$ 0.35\% peak value. The two
  images and lens galaxy are labelled as A, B and G respectively.}
\label{fig:hst}
\end{figure}

To locate the lens galaxy and hence derive the likely position of the
central image, we used the HST WFPC2 I-band image (Fig.~\ref{fig:hst})
of B1030+074 \citep{xanthopoulos.98.mn} from the HST archive to
measure the position of the lens galaxy. We can clearly see both the
lensed images and the lens galaxy which has an extended structure to
the west.  The separation from lens galaxy G to the B image is
$127\pm10$ mas in position angle $149^\circ.7\pm4^\circ.0$ from north
through east; there is no obvious detection of a radio source above
the local noise level at the position of G.

The flux density of both A and B components has varied ($\sim 10\%$)
over the seven-year span of observations. Variation was also seen in a
monitoring campaign reported by \citet{xanthopoulos.00.iau} which
consisted of 47 epochs of VLA observations spread over 240 days. Both
datasets are not inconsistent with intrinsic variation of the source,
together with an expected time delay of $\sim$ 100--150 days predicted
by simple lens models. In the case of J1838$-$3427, \citet{winn.04.aj}
deduce that Galactic scintillation makes a major contribution to the
variability; this process is less likely to operate in B1030+074 due
to the relatively high ($b=52^{\circ}$) Galactic latitude.

\section{Derivation of the upper limit}\label{noise}

The 3-$\sigma$ limit of our final HSA map is about 70~\mujybm, and this
level has traditionally been regarded as the effective upper limit on
the central-image flux density if there is no obvious detection.  This
is not the case, for two reasons.  First, the image has undergone a
{\sc clean}ing and self-calibration process, which may have removed
flux from any central image. Second, the uncertainty in the position
of the central image means that the image statistics must be
considered carefully. We consider each issue in turn.

\subsection{{\sc clean}ing and self-calibration effects}

The use of the {\sc clean} algorithm in the
mapping procedure may result in elimination of faint images of a few
$\sigma$, a problem known as ``{\sc clean} bias'' \citep{condon.98.aj}. 
Simulations with artificial sources show that sources up to 6$\sigma$ 
may be eliminated in this way if excessively {\sc clean}ed, but that 
use of {\sc clean} boxes around the true source position should allow 
them to be recovered. It is also possible that the use of
self-calibration may lead to the adjustment of the data in such a
way as to eliminate weak sources, by partially absorbing them into 
the calculated gains for each telescope.

\begin{figure*}
\begin{tabular}{c}
\psfig{figure=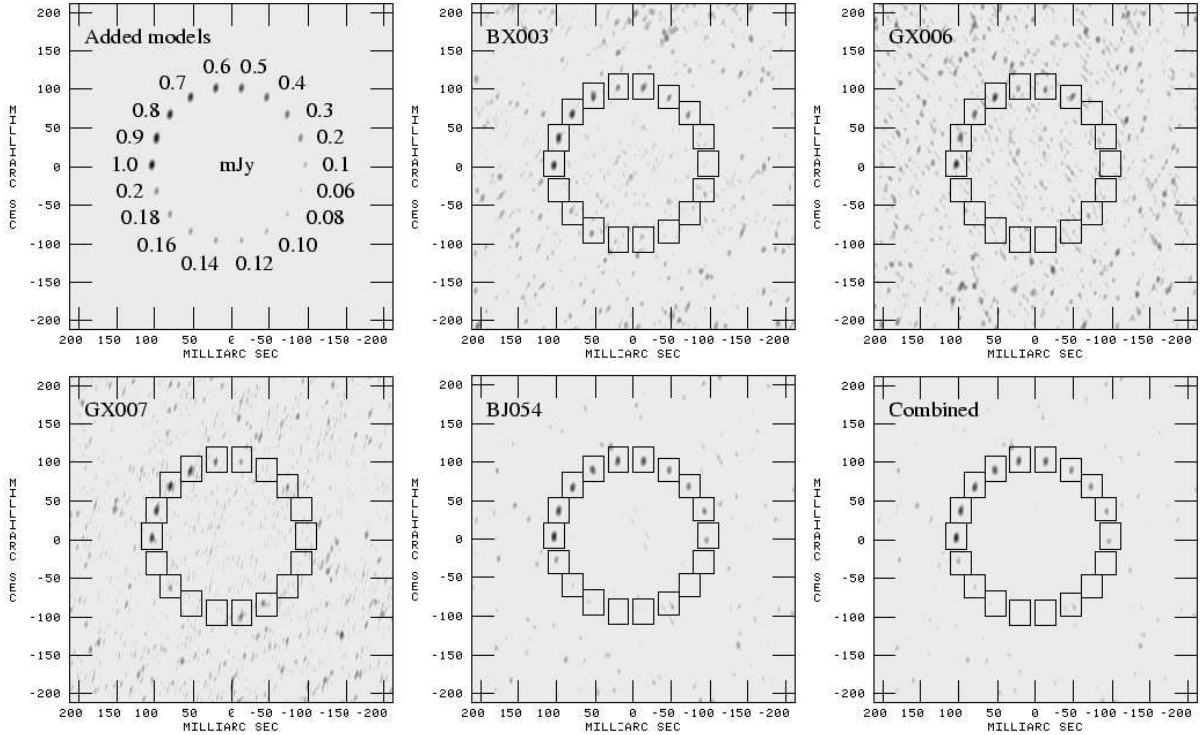,width=16cm}
\end{tabular}
\caption{ Artificial source `ring' recovery test. The solid rectangles
  are used as a guide to the source positions. The positions of our 18
  added models in the $uv$ data form a circle with $20^\circ$
  separation from each other. The upper half-circle includes 10
  sources from 1000~$\mu$Jy to 100~$\mu$Jy in steps of 100~$\mu$Jy,
  and the lower half-circle includes 8 sources from 200~$\mu$Jy to
  60~$\mu$Jy in steps of 20~$\mu$Jy. For comparison, all maps are set
  to linear scale with a flux density range of from 0.03~mJy to
  1~mJy.}
\label{fig:rings}
\end{figure*}

In order to assess this problem, we have used the {\sc aips} task {\sc
uvmod} to add artificial point sources with a range of fluxes to the
calibrated $uv$ data, and determined whether they are recovered during
passage through our self-calibration and mapping procedure. 18
artificial point sources were used simultaneously, arranged in a ring
of diameter 200~mas (see Fig.~\ref{fig:rings}) and with flux densities
ranging from 60~$\mu$Jy to 1~mJy.  This also allowed us the
opportunity to adjust the procedure in order to recover the weakest
possible source. The philosophy throughout the process has been to
apply the same procedure to the artificial images as to the possible
central image.

For reasons which we have been unable to determine, the use of
relatively tight {\sc clean} boxes around components A and B gives
poor noise characteristics and a high rms in the rest of the image,
including the position of the possible central image. On the other
hand, {\sc clean}ing the whole image gives the low noise (24~\mujybm)
seen in Fig.~\ref{fig:whole} but relatively poor recovery of the artificial
images. An alternative boxing scheme is to {\sc clean} the majority of
the image, leaving a gap around the ring of artificial sources and the
position of the central image, but in addition including a small box
of 10$\times$10~mas$^2$ around the expected position of each individual
artificial image and of the central image. This gives good results in
that artificial sources of 100~$\mu$Jy can be recovered; because the
artificial sources and any potential central image are treated
identically, we can be confident that any real central image at this
flux density level could be detected.

\begin{figure}
\begin{tabular}{c}
\psfig{figure=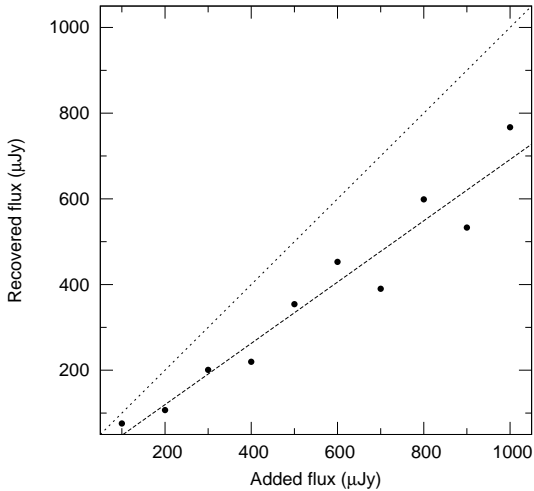,angle=-90,width=7cm}
\end{tabular}
\caption{ Flux loss in the HSA map from our mapping procedure,
expressed in a plot of the flux density of an artificially added
source against the recovered flux density. The dashed line is a linear
fit to the data which has a slope of 0.72. The dotted diagonal line
has a slope of 1. }
\label{fig:fluxloss}
\end{figure}

The results are shown in Figs.~\ref{fig:rings} and
\ref{fig:fluxloss}. The best recovery is from our HSA data, with
recovery of 100~$\mu$Jy sources. However, in the previous three VLBI
observations, stronger sources than this (up to 200~$\mu$Jy) were in
many cases not recovered.  In both cases, the recovery of artificial
test sources is only achieved at between 5 to 10 times the expected
thermal noise. In addition, our procedure does not recover the full
flux density of the artificial sources.  In Fig.~\ref{fig:fluxloss} we
plot the flux density inserted against the flux density
recovered. Although there is considerable scatter in the relation, a
linear fit gives a gradient of 0.72, implying a $\sim 30\%$ reduction
in flux density caused by the self-calibration and mapping process.

%

%

One possibility is that the weighting of the individual antennas could
be responsible for the non-recovery of weak sources, since the HSA is
a combination of ten small antennas and three (Green Bank Telescope,
the phased VLA and Arecibo) which have orders of magnitude greater
collecting area. If the large telescopes are given their full
statistical weight, the number of constraints used to derive the
amplitude and phase corrections applied by self-calibration is
effectively reduced, because few baselines are weighted significantly
and in fact the signal is dominated by three highly sensitive
baselines. This in turn means that the solution for the amplitude and
phase errors on the large telescopes is under-constrained, and allows
genuine structure to be removed by the corrections. To test this
possibility, we weighted down the large telescopes, up to the extreme
case of treating all telescopes with equal weight, at a cost of an
increase in the thermal noise. However, we did not find much
improvement of the recovery of the added sources, and in fact most
added sources suffered worse flux loss.

\subsection{Statistics of non-detections}

We will take a simple and conservative approach to assessing the
detectability of a third image in our data. We would like to be able to
give a value for probability that we would have detected a third image
of flux density $\geq S$ in our data, if it were there.  Unfortunately
we cannot do this. We can assess the probability of detecting a source
of flux density $S$ by doing simulations but not the cumulative
probability of detecting an image $\geq S$. This is because we do not
know the expected distribution of the third image flux densities over
which we should integrate. However, one thing we can be confident of
is that the detectability of images increases with increasing flux
density. This means that, if we assess the probability $P^{*}$ of
detecting images of flux density $S$ by adding artificial sources of
this flux density to our observed data, and see how many we recover,
the result can give us a lower limit on the desired cumulative
probability. We will adopt $P^{*} = 0.9$ as our conservative limit on
the probability of detecting an image, if it had a strength of $\geq
S$. We now describe how we search the dataset before we return to the
probability calculation.

Because of uncertainties in the galaxy position, and because the
third-image position depends on the adoption of a model, we have to
consider the pixel values in the final map over a search box rather
than at a single position.

\begin{figure}
\begin{tabular}{c}
\psfig{figure=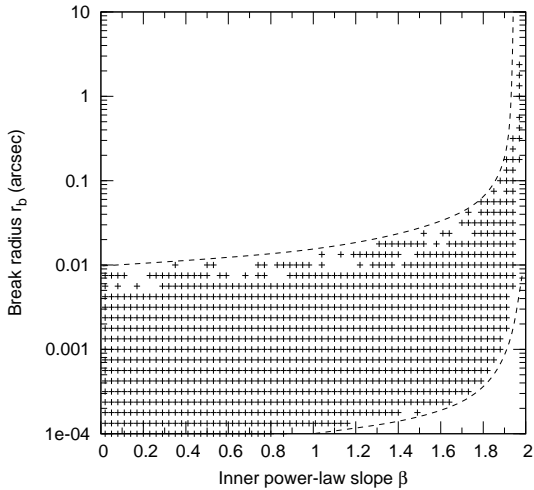,width=7cm}
\end{tabular}
\caption{ Grid-search sampling in $\beta-r_b$ plane. The `+' symbol
  denotes the grid points we sampled to determine the predicted
  central image position. Above the dotted line, the central image is
  predicted at a level which would be obvious from inspection of the
  radio map. }
\label{fig:region}
\end{figure}

\begin{figure*}
\begin{tabular}{c}
\psfig{figure=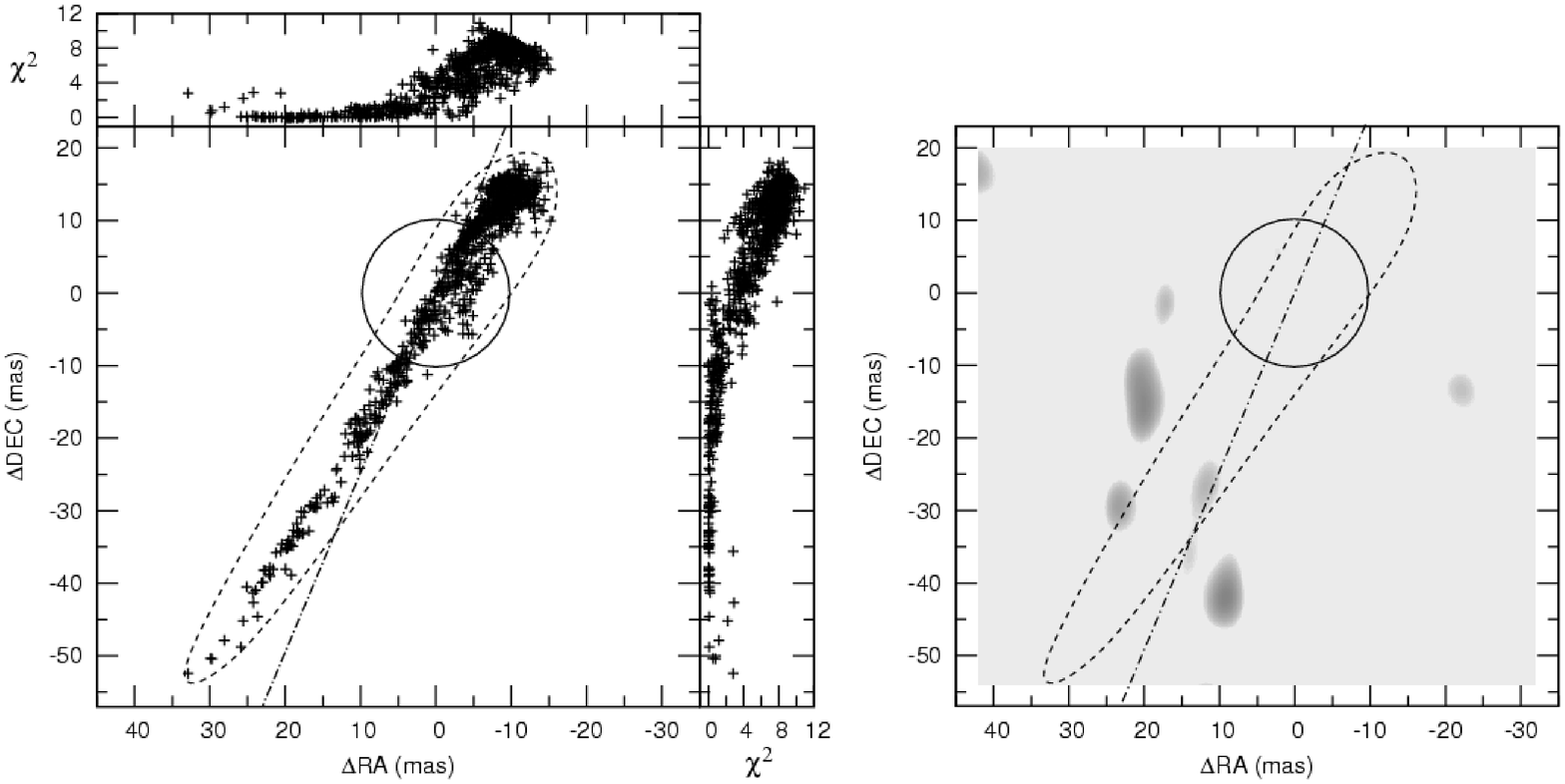,width=14cm}
\end{tabular}
\caption{ Predicted third image positions from cusped models. The
  solid circle is the galaxy position with $1\sigma$ error. The
  dot-dashed line denotes the major axis of the elliptical mass
  distribution. In the left figure, the `+' symbols denote the
  predicted third image positions in the image plane from each fit
  using different values of $r_b$ and $\beta$. The club-shaped dashed
  line pattern encloses all the predicted third image positions in our
  test and has an area of 840~mas$^2$; the top and right panels show
  the $\chi^2$ distributions of each fit, as a function of $\Delta$RA
  and $\Delta$DEC respectively. In the right figure, we show the HSA
  map of the predicted third image region overlapped with the club
  pattern, and the rms uncertainty of the galaxy position as a
  circle. In these models, the galaxy position has been allowed to
  vary, but with a Gaussian penalty function based on our knowledge of
  its position from the HST image and the rms uncertainty.}
\label{fig:club}
\end{figure*}

Defining the search box depends on the adoption of a model. We defer
detailed discussion of this model to the next section, but note that
its two principal free parameters are the break radius $r_b$ and inner
power-law slope $\beta$ of the galaxy mass distribution.
Fig.~\ref{fig:region} shows the parameter space explored to find possible
positions of the central image, and the predicted third image
positions are shown in Fig.~\ref{fig:club}. From Fig.~\ref{fig:club},
we found that all the possible third image positions are enclosed in a
club-shaped pattern along the major axis of the elliptical mass
distribution. The area of the club shape is about 840 mas$^2$, i.e.,
of about 24 beams.

Using 10$\times$10 mas$^2$ {\sc clean} boxes in this area for
imaging, as described in the previous subsection, gives a maximum flux
density within this region of 42~$\mu$Jy. This value is consistent
with being a noise spike, and in fact if the noise were Gaussian
(which it is not) one would expect several such values within the
840-mas$^2$ search box.

We now return to the probability calculation. We recover no source
stronger than 42~$\mu$Jy from the search box. What is the flux density,
$S$, that an image would have to have for $P^{*} = 0.9$? To find this
we have experimented with artificial sources of different flux
densities and find that using sources of 180~$\mu$Jy, 90\% of them are
recovered with flux densities greater than 42~$\mu$Jy. Therefore our
conclusion is that the probability of detecting an image using our
search technique is $\geq 90\%$, provided its flux density is $\geq
180~\mu$Jy. A more detailed interpretation of this upper limit is
given in Appendix~\ref{app:upp}.

%

%

\section{Lens modelling}\label{model}

The mass distribution of galaxies has been a subject of debate in the
last two decades. On the scale of the outer images in lens systems
(normally a few kiloparsecs) an isothermal profile is usually a good
fit \citep[e.g.][]{koopmans.06.apj}. In the past, power-law models with
a nearly isothermal profile and a central flat core were used
\citep[e.g.][]{lauer.85.apj}. However, numerical simulations
\citep{navarro.97.apj} and HST observations \citep{faber.97.aj} have
been used to argue for a more general description in which galaxies
have a two-power-law form with a central cusp. Normally, for a weak
cusped model, the odd number theorem applies and a third image exists
\citep{evans.98.mn}.

We adopt a typical spherical density profile of a cusped model
\citep[e.g.][]{munoz.01.apj}:
\begin{equation}
\rho(r)={\rho_0\over(r/r_b)^\beta[1+(r/r_b)^2]^{(\eta-\beta)/2}}
\end{equation}
where $\rho$ is the mass density, $\rho_0$ is the scale factor, $r$ is
the radius, $\eta$ is the outer slope, $\beta$ is the inner slope, and
$r_b$ is the break radius. This two power-law distribution degenerates
to single power-law distributions $r^{-\eta}$ or $r^{-\beta}$
respectively when $r\gg r_b$ or $r\ll r_b$. In the limit
$\beta\rightarrow 0$, we have a cored model, whereas in the limit
$\beta=\eta=2.0$, we recover a singular isothermal model (SIE). More
generally, for the ellipsoidal models, the $r$ should be replaced by
triaxial coordinates ($r^2=x^2/a^2+y^2/b^2+z^2/c^2$)
\citep{chae.98.apj}.

Numerous authors have pointed out that regions of the $r_b$--$\beta$
parameter space can be ruled out by the non-detection of central
images in gravitational lens systems. In all cases a power-law model
with an isothermal slope or steeper causes infinite demagnification of
the central image. Early studies by \citet*{wallington.93.apj}
examined softened power-law models with a core; in this case an upper
limit on the core radius could be derived. \citet*{evans.02.apj}
developed a general formalism for cored profiles and found that for
observed systems with missing central images, the typical core radius
of the lensing galaxy must be $<$300~pc; they concluded that the mass
profile must be nearly cusped and that the cusp must be isothermal or
stronger. \citet*{rusin.01.apj} used the non-detection of central
images in a number of Cosmic Lens All-Sky Survey (CLASS)
\citep{myers.03.mn,browne.03.mn} gravitational lens systems to infer
limits on single power-law models (in which $\beta=\eta$ and $r_b$ is
therefore irrelevant) and for B1030+074 found that the then existing
limit on the central image implied a central power law slope
$\beta>1.91$.  \citet{munoz.01.apj}, \citet{winn.03.apj, winn.04.aj}
and \citet{boyce.06.apj} explicitly studied the allowed region of the
$r_b$--$\beta$ plane for various lens systems, assuming an isothermal
outer slope ($\eta$=2). In each case the non-detection of a central
image allows the exclusion of a region of this plane with a relatively
flat inner slope (low $\beta$) and large break radius (high $r_b$),
and this restriction becomes more powerful as the limit on the ratio
between the central image and the brightest image becomes more
stringent. In the case of a detection \citep{winn.04.nat} an allowed
locus on this plane can be inferred.

\begin{figure}
\begin{tabular}{cc}
\psfig{figure=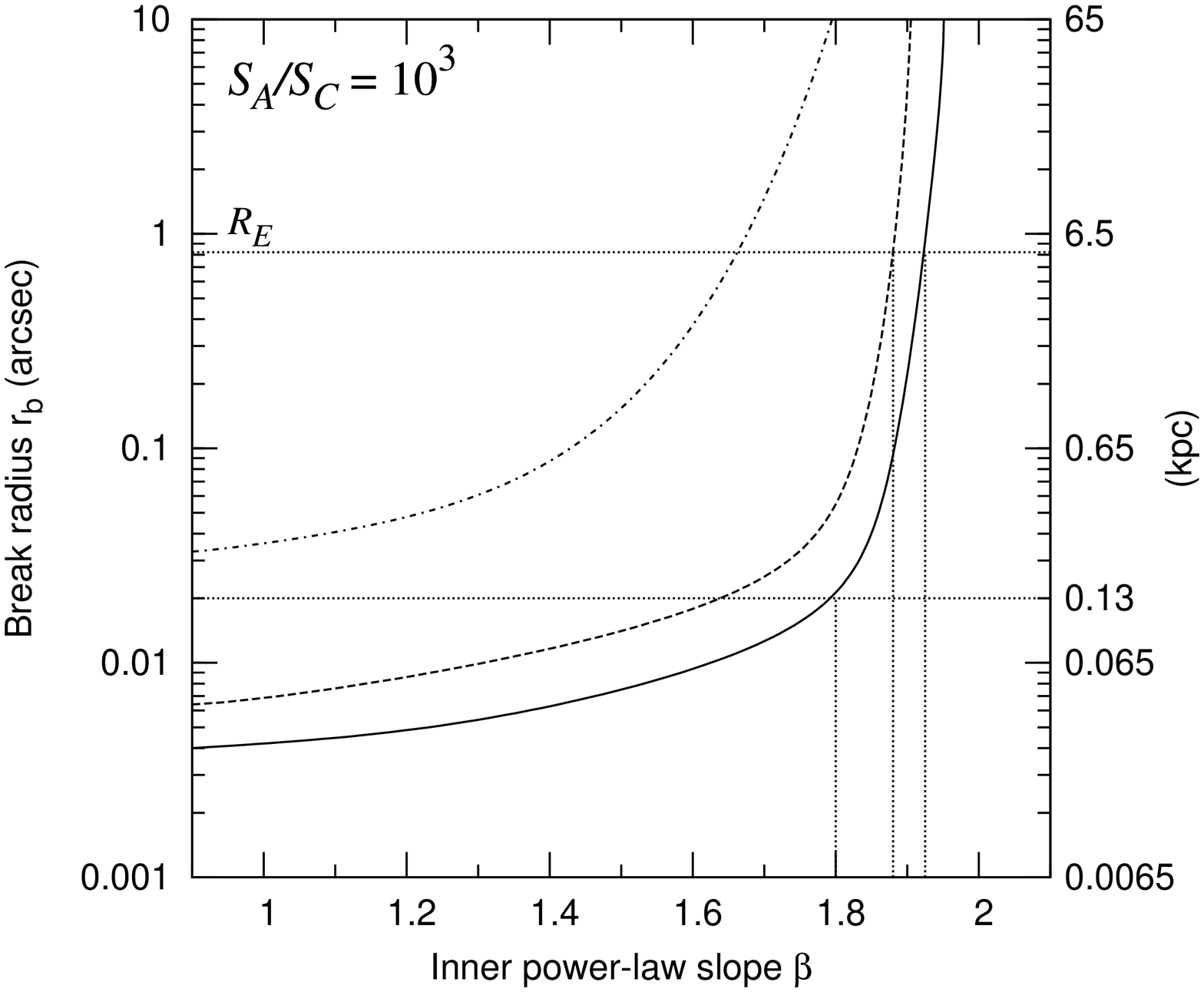,angle=-90,width=8cm} \\
\psfig{figure=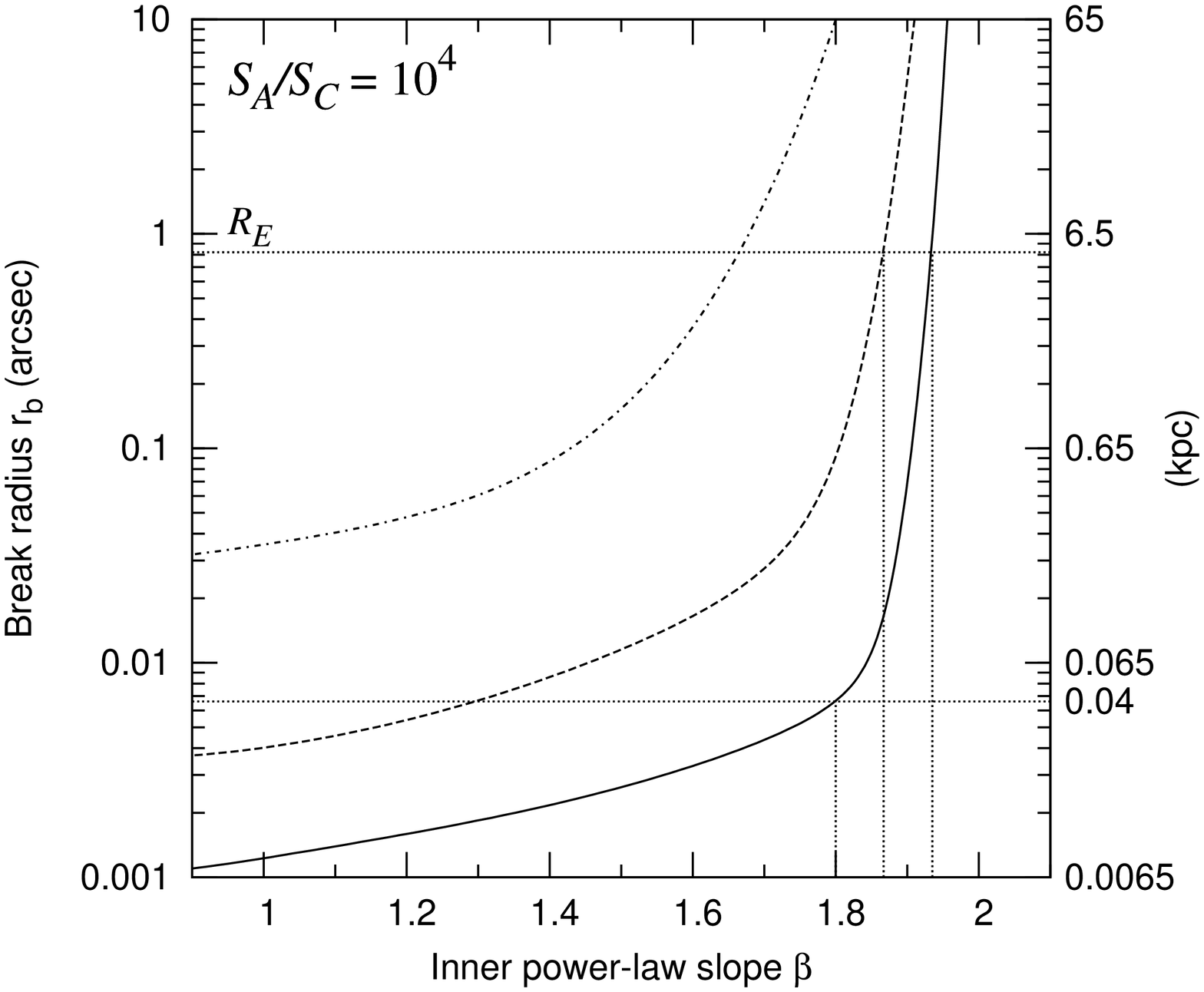,angle=-90,width=8cm} \\
\psfig{figure=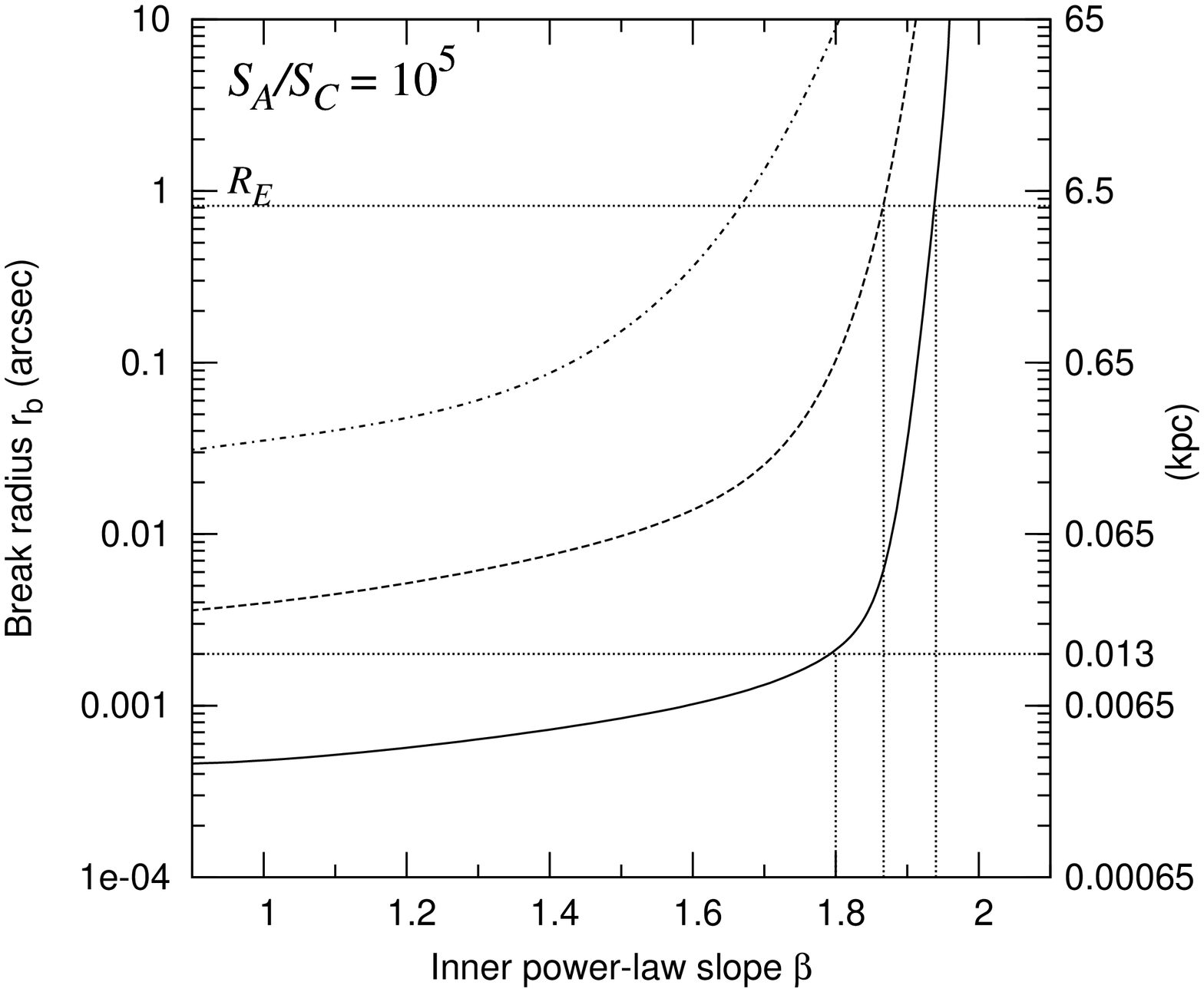,angle=-90,width=8cm}
\end{tabular}
\caption{Third image detection regions in break radius - inner
  power-law slope diagram. From the top to the middle and the bottom,
  the three figures are calculated with primary/central image
  magnification ratios of $10^3$, $10^4$ and $10^5$ respectively. The
  solid lines are the results calculated by the cusped model, the
  dashed lines are the results calculated by the cusped model with a
  $2.5\times10^8~\rm M_\odot$ black hole, the dot-dashed lines are the
  results calculated by the cusped model with a $2.5\times10^9~\rm
  M_\odot$ black hole. $R_E$ is the major axis from a fitted SIE
  model.}
\label{fig:breakslop}
\end{figure}

The number of degrees of freedom in a simple cusp model is negative,
since the free parameters (dimensionless surface density $\kappa_0$,
ellipticity $e$, position angle $\theta$, $r_b$, $\beta$, source and
galaxy ($x,y$) positions) outnumber the measured parameters
\citep[e.g.][]{boyce.06.apj}.  Following \citet{winn.03.apj} and
\citet{boyce.06.apj}, we therefore investigated the behaviour of
cusped models for B1030+074 with different values of inner power-law
slope, $\beta$, and break radius, $r_{b}$, together with a fixed
isothermal outer power-law slope $\eta=2$. This was repeated for
different flux ratios between the primary image A and the third image
C (see Fig.~\ref{fig:breakslop}). We used the fast Fourier expansion
method of \citet{chae.98.apj} and \citet{chae.02.apj} for this
purpose.  The higher the observational limit on the magnification
ratio, the larger the non-detection region above the solid lines. For
our value of $S_A/S_C \sim 1000$ (from our derived upper limit on
$S_C$) and $r_b\gtrsim130$ pc, we obtain an inner power-law slope
$\beta$ greater than 1.8.

%

One possibility is that a massive central black hole in the lens
galaxy could suppress the central image
\citep{mao.01.mn,keeton.01.apj,keeton.03.apj}. \citet{keeton.03.apj}
analysed the likely suppression of central images for different black
hole masses and concluded that the average degree of suppression is
quite small; the exception is in those cases where the central image
is extremely faint in any case. \citet{rusin.01.apj} calculate
explicitly the case of B1030+074, using a single power-law model
($\beta=\eta$) together with a plausible black hole mass, and conclude
that the limit on the power-law slope needs to be relaxed from
$\beta>1.91$ to $\beta>1.83$.

A plausible black hole mass can be estimated by using the empirical
correlation between the central black hole mass and the galactic bulge
velocity dispersion
\citep{ferrarese.00.apj,gebhardt.00.apj,tremaine.02.apj}
\begin{equation}
\log(M_{BH}/M_{\odot})=\alpha+\beta \log(\sigma /\sigma_0),
\end{equation}
where $\beta=4.02\pm0.32$, $\alpha=8.13\pm0.06$,
$\sigma_0=200{\rm ~kms^{-1}}$ are the fitted
values from \citet{tremaine.02.apj}. Since the galactic mass
distribution is very close to isothermal, we can use the relation
\begin{equation}
\Delta\theta/2=4\pi(\sigma^2/c^2)(D_{ds}/D_s)
\end{equation}
to estimate the velocity dispersion of the lens galaxy, where
$\Delta\theta$ is the angular separation between image A and B,
$D_{ds}$ and $D_s$ are the angular diameter distances between lens and
source, and source and observer, respectively. For the case of
B1030+074, $\Delta\theta\simeq1.567$ arcsec, which gives a velocity
dispersion $\sigma$ of 233 $\rm kms^{-1}$, and hence a central black
hole mass $M_{BH}$ of about $2.5\times10^8~\rm M_\odot$\footnote{Here
and elsewhere in this paper we use a flat $\Lambda$CDM cosmological
model with $\Omega_m=0.3$, $\Omega_\Lambda=0.7$ and $H_0=72$
kms$^{-1}$Mpc$^{-1}$.}.  Applying the cusped model together with this
black hole mass, we can get the modified relation between break radius
$r_b$ and inner power-law slope $\beta$, which is shown in
Fig.~\ref{fig:breakslop}. In the limit of large break radius
($r_b\rightarrow\infty$), the cusped model approaches a single
power-law model similar to the simulation of \citet{rusin.01.apj}
previously described. In our model, the asymptotic value of inner
power-law slope $\beta\simeq1.95$, and $\beta\simeq1.91$ with a
$2.5\times10^8~\rm M_\odot$ central black hole in the lensing
galaxy. If we cut off the break radius at the Einstein radius of an
isothermal model, we get an inner power-law slope $\beta\simeq1.93$,
and $\beta\simeq1.88$ with a $2.5\times10^8~\rm M_\odot$ central black
hole. Obviously, a central massive black hole affects the $r_b-\beta$
curve more when the flux ratio $S_A/S_C$ is bigger. The $2.5\times10^8~\rm M_\odot$ black hole derived from the $M_{BH}-\sigma$ relation does
not soften the cusped mass profile significantly. We would need to add
a black hole of 10 times greater mass in order for the cusped mass
profile to be significantly shallower for the same $S_A/S_C$
constraint. This result agrees well with Keeton's conclusion from
simulations of star+halo CDM mass models, that black holes would have
to lie off the $M_{BH}\sim\sigma$ relation by at least a factor of 10
in mass to significantly affect the central images
\citep{keeton.01.apj}.


\section{Superluminal jet?} \label{superlum}

\begin{figure}
\begin{tabular}{cc}
\psfig{figure=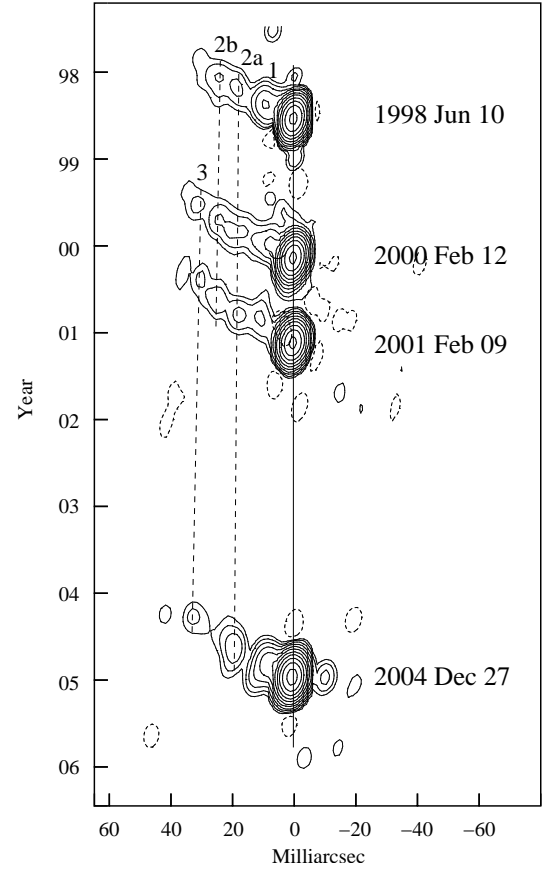,width=7cm} \\
\psfig{figure=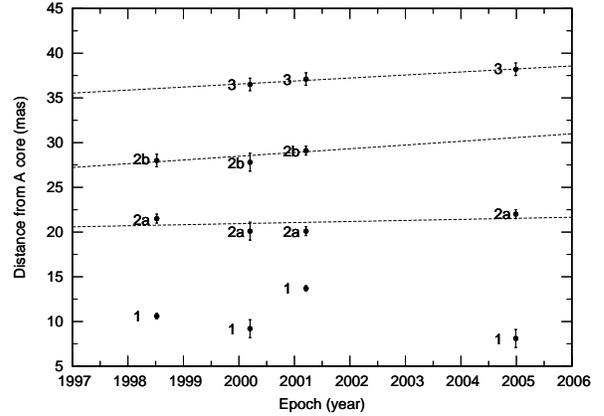,angle=-90,width=8cm}
\end{tabular}
\caption{Jet proper motion of B1030+074 image A. The top figure shows
  the most likely correspondence of outgoing individual jet
  components. All images are restored with the same restoring beam and
  aligned assuming a stationary core position where the vertical scale
  is proportional to the time intervals between the different
  observations. The bottom figure shows the distances between jet
  components and the core at a position angle of $60^\circ.1$. The
  labelled numbers show the correspondence of jet components. The
  three dashed lines indicate the results from weighted fits to proper
  motions of the jet components (2a, 2b, 3), corresponding to the
  three dashed outlines in the top figure. Fitted slopes are
  $0.34\pm0.20$~mas~yr$^{-1}$ for component 3 ($\chi^2=0.07$),
  $0.42\pm0.32$~mas~yr$^{-1}$ for component 2a ($\chi^2=0.7$) and
  $0.12\pm0.11$~mas~yr$^{-1}$ ($\chi^2=7.6$) for component 2b.}
\label{fig:jetmov}
\end{figure}

The jet within the A component is potentially useful
astrophysically. First, because of the stretching induced by lensing
magnification, any superluminal motion is easier to observe.
Superluminal motion was predicted four decades ago
\citep{rees.66.nat}, and many superluminal jets have been discovered
since then \citep{vermeulen.94.apj,britzen.99.asp}. These superluminal
motions are typically a few times the speed of light, but would appear
larger in the image plane because of magnification; in extreme
high-magnification cases speeds up to 100~c would in principle be
observable \citep*{hogg.94.mn}. The nearly seven-year span of our
observations of B1030+074 gives us a good opportunity to investigate
the existence of any superluminal movement of the jet in a lens system
(see Fig.~\ref{fig:jetmov}). Second, any components found in both A
and B give extra information to constrain the mass model. The A image
in the B1030+074 lens system has a very significant jet while the jet
is not resolved in the B image.

%

Maps of image A at all four epochs, made using the same {\sc clean}
restoring beam, are shown in Fig.~\ref{fig:jetmov}.  At each epoch the
jet exhibits a number of knots. Although there are always ambiguities
in identifying such jet components from epoch to epoch, the most
conservative correspondence is shown by the dotted lines in
Fig.~\ref{fig:jetmov}. There is a suggestion of outward motion of the
outermost jet knot 3; the fitted slope implies a proper motion of
about $0.34\pm0.20$~mas~yr$^{-1}$, corresponding to a velocity of the
jet component in the image plane of about $7.2\pm4.2$~c. The fitted
magnification of the A image for an SIE model is about $\mu\simeq$ 3.0
in the direction along the jet, so in the source plane the jet proper
motion is about $3.0\pm1.7$~c. From top to bottom, the other two jet
components give proper motions in the source plane of $3.7\pm2.8$~c
and $1.1\pm0.9$~c, respectively. The determination of the true jet
proper motion relies on the magnification factor which is model
dependent. The cusp model can give a range of A image magnification
from 3.0 to 5.0.  With the $S_A/S_C\sim1000$ non-detection constraint
the fitted cusped model gives a typical A image magnification of about
2.8, and 3.1 with a $2.5\times10^8~\rm M_\odot$ black hole estimated
from the $M_{BH}-\sigma$ relation. A larger magnification of image A
could be achieved if we add a bigger black hole mass to suppress the
central image.

\begin{figure}
\begin{tabular}{cc}
\psfig{figure=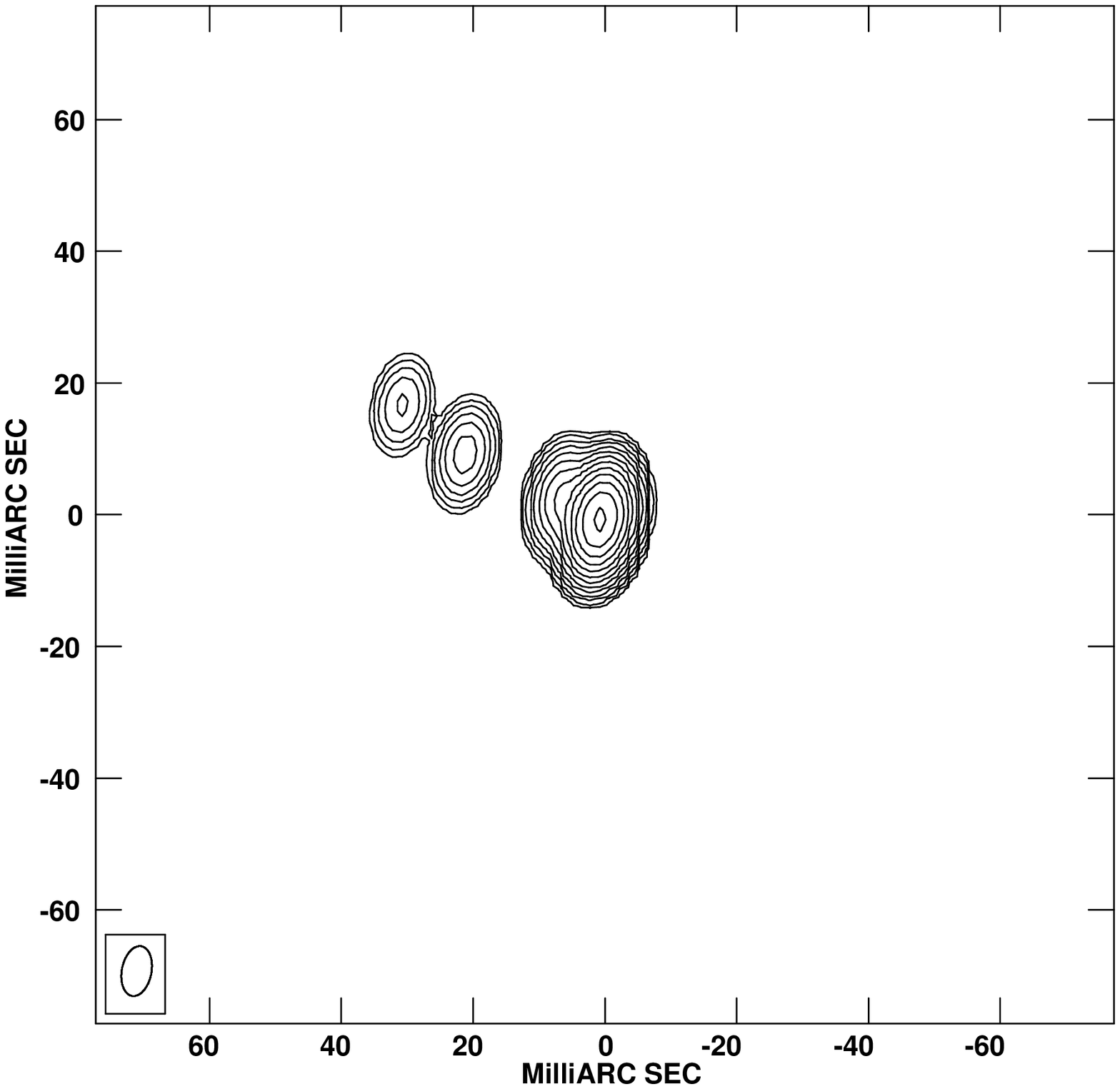,width=4cm}
\psfig{figure=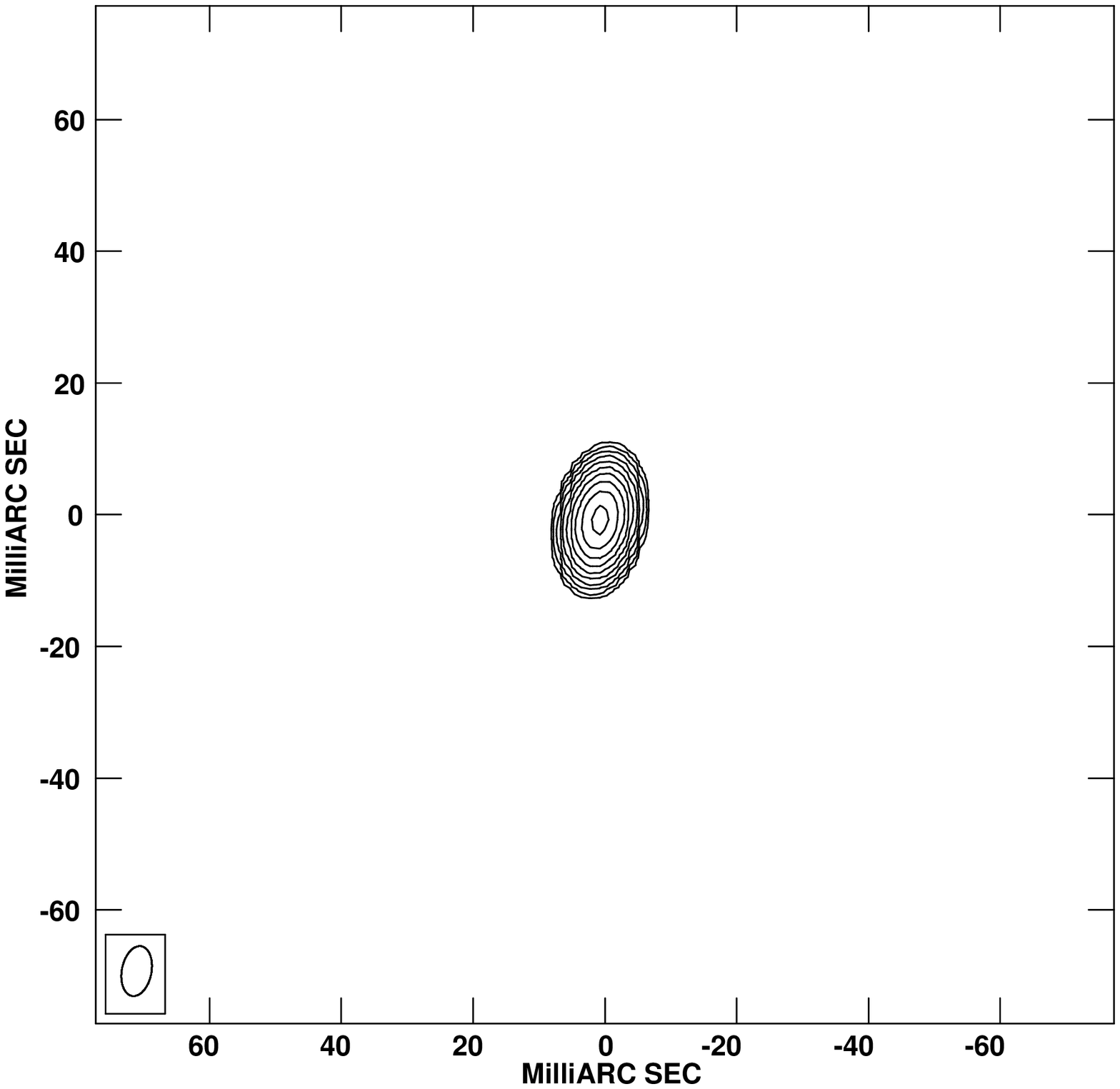,width=4cm} \\
\psfig{figure=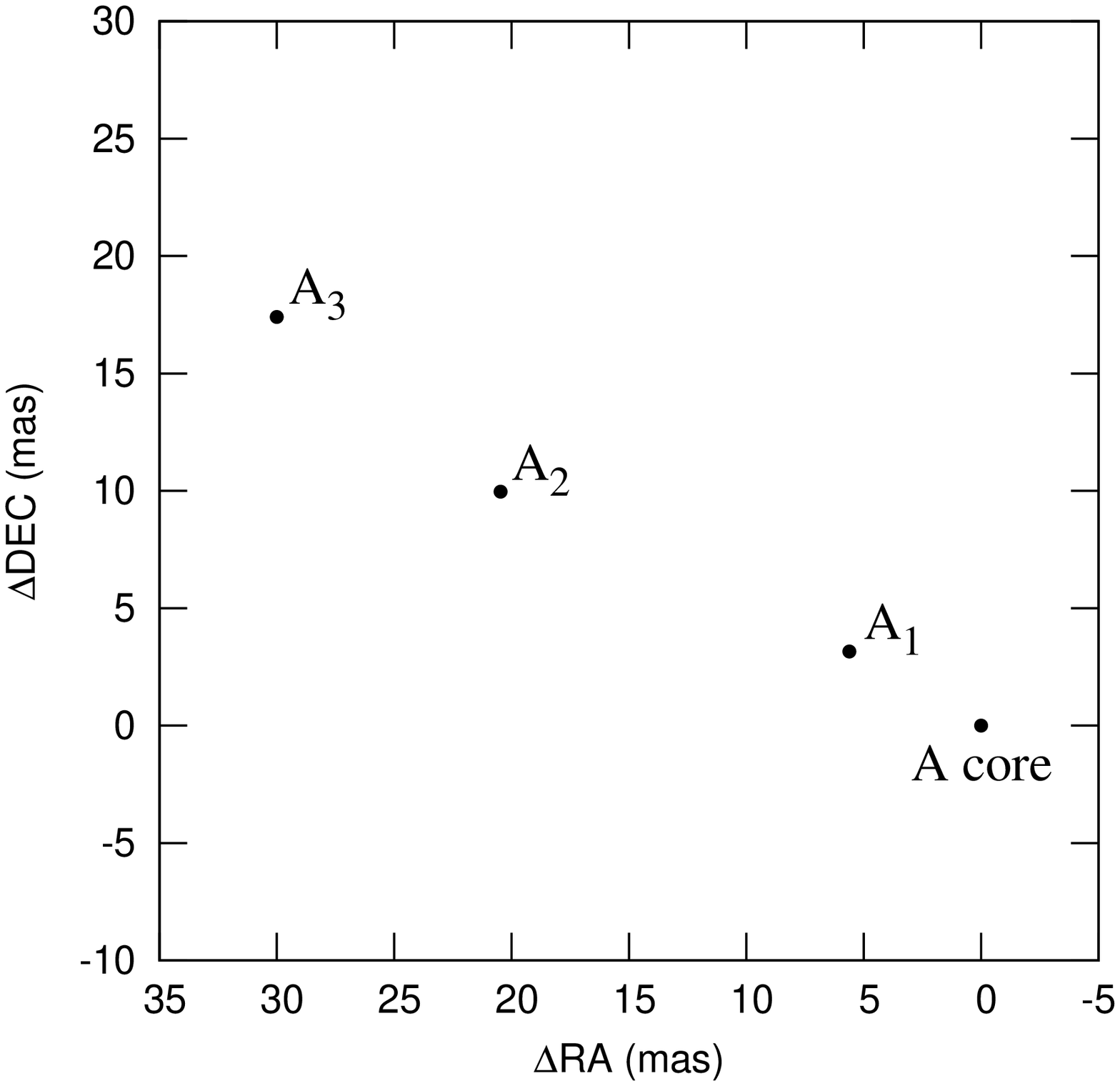,angle=-90,height=4cm}
\psfig{figure=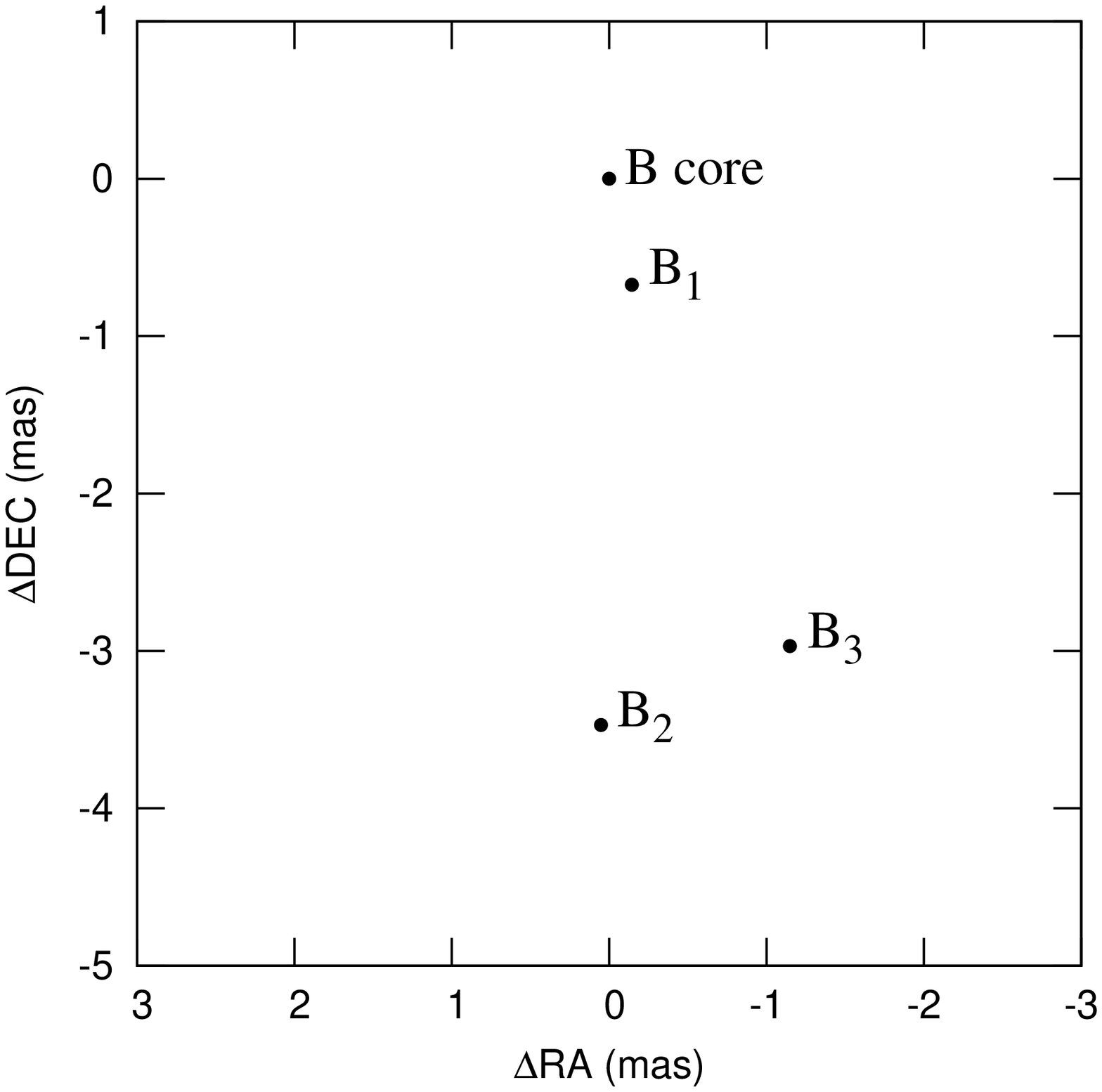,angle=-90,height=4cm} \\
\end{tabular}
\caption{ Simulated images from the fitted SIE model. They are
convolved with a restoring beam of $7.7\times4.5$ mas$^2$ at a
position angle of $-10^\circ.85$. The contour levels here are the same
as Fig.~\ref{fig:whole}. The jet structure in image A is detectable
while image B still looks point-like.}
\label{fig:simu}
\end{figure}

The jet in the B image is not resolved. Using the components of the
observed A jet from the VLBI/HSA image together with the fitted lens
model, we simulated the structure of the B jet using a point spread
function (PSF) similar to the restoring beam in our VLBI images
(Fig.~\ref{fig:simu}). The simulation confirms that we would not
expect to resolve the B jet structure using our current observations.

%

\section{Conclusions}

We have presented a new HSA 1.7-GHz observation of the double-image
lens system B1030+074 which was aimed at finding a third image to help
to understand the central mass distribution of the lens
galaxy. Careful consideration of observational and statistical effects
shows that the 90\% confidence level of the non-detection in such
cases is about 7 times the rms noise (in this case corresponding to
180~$\mu$Jy).  This gives a minimum flux ratio of about 10$^{3}$
between the brightest image and the central image. By comparing with
predictions from a cusped model \citep{munoz.01.apj}, this
non-detection constrains the break radius - inner power-law
($r_b-\beta$) relation, which excludes most of the $r_b-\beta$ region
depicted in Fig.~\ref{fig:club} and implies a steep inner power-law
slope close to isothermal. Following the $M_{BH}-\sigma$ relation
\citep{tremaine.02.apj}, we predicted a central black hole mass of
about $2.5\times10^8~\rm M_\odot$.  However, including this black hole
mass into the cusped model did not affect the inner power-law slope
very much, and would only do so if the black hole mass were increased
by a factor 10. The result is consistent with simulations from CDM
mass distributions of the lens galaxies \citep{keeton.03.apj}.

Four-epoch observations gave a possible detection of superluminal
motion in the jet. The determination of the true superluminal velocity
also depends on the mass model, but with the fitted cusped model, the
superluminal motion of the jet is about 3~c in the source plane,
although subject to large errors at this stage.

The current 1.7-GHz HSA observation cannot resolve the jet in the B
image even with a super-resolved restoring beam. We simulated the B
jet from the fitted mass model and convolved it with the restoring
beam. This simulated B image still looks like a point
source, which is consistent with the observation. A 5-GHz observation
may possibly resolve it if the spectral index is not so steep as to
make it below the surface brightness detection limit.

In the next few years, interferometer arrays will come online which
should make the detection of central images in gravitational lens
systems routine. In particular, the e-MERLIN array will have an L-band
sensitivity of 4.0~\mujybm ~with 12~hours on-source integration. With a
resolution of 50~mas at 5~GHz, this instrument is especially
well-matched to the problem, since the simpler image structures are
more easily mapped with high dynamic range, and the region to be
searched to find the central image is at most a few resolution
elements. If current theoretical work is correct
\citep[e.g.][]{keeton.03.apj} this should allow detection of central
images, and therefore greatly improved constraints on mass models, in
many known radio lens systems.

\subsection*{Acknowledgements} 
This work was carried out under financial support from a Marie Curie
Research Training Network - Astrophysics Network of Galaxy LEnsing
Studies (`ANGLES' MRTN-CT-2003-505183). The authors are grateful to
Edward Boyce, Shude Mao, John McKean and Roger Barlow for
discussions. We also thank the referee Joshua Winn for his constructive
comments. The VLA, VLBA and GBT are operated by the National Radio
Astronomy Observatory which is a facility of the National Science
Foundation operated under cooperative agreement by Associated
Universities Inc. The Arecibo Observatory is part of the National
Astronomy and Ionosphere Center, which is operated by Cornell
University under a cooperative agreement with the National Science
Foundation.  This research is based in part on observations with the
Hubble Space Telescope obtained at the Space Telescope Science
Institute, which is operated by Associated Universities for Research
in Astronomy for NASA under NASA grant no. NAS5-26555.

\appendix

\makeatletter
\def\ExtendSymbol#1#2#3#4#5{\ext@arrow 0099{\arrowfill@#1#2#3}{#4}{#5}}
\def\RightExtendSymbol#1#2#3#4#5{\ext@arrow 0359{\arrowfill@#1#2#3}{#4}{#5}}
\def\LeftExtendSymbol#1#2#3#4#5{\ext@arrow 6095{\arrowfill@#1#2#3}{#4}{#5}}
\makeatother
\newcommand\longeq[2][]{\ExtendSymbol{=}{=}{=}{#1}{#2}}

\section{about the upper limit}\label{app:upp}

There are two schools of inferential statistics: frequentist and
Bayesian. Each gives different interpretation of probability as
well as confidence level. We will investigate two different views on
the upper limit. 

In the following discussion, we use the following notations: the
conditional probability $P(S|D)$ is for ``getting the source, given
the data'', and $P(D|S)$ for ``getting the data, given the source'';
$P(D)$ and $P(S)$ are for data and source probability distributions
respectively.

Our specification of the upper limit with desired confidence level is
directly from the standard {\em Neyman construction}
\citep{neyman.37.rspta}. To set an upper limit on our
non-detection, we simulated the input with artificial sources and
analysed the output from our mapping procedure.

Our test is based on several factors:

$\diamond$ The noise in a {\sc clean}ed map is not an objective
statistic. Though we can manage to {\sc clean} it down to thermal
noise, it is pointless if the expected weak source is {\sc clean}ed
out.

$\diamond$ The recovery of an injected artificial source is stationary
and related to its input position and strength, i.e., if we repeat
trials with the same input position and strength, the recovery of the
artificial source through the same mapping procedure is same, either
false dismissal or not.

$\diamond$ The higher the input strength, the better recovery we get,
while the lower the input strength the worse recovery. For a source
above 250 $\mu$Jy, the recovery is almost certain though it will
definitely suffer a lossy {\sc clean}ing; below 100 $\mu$Jy, the
recovery is hardly distinguishable from noise.

$\diamond$ The background noise of a {\sc clean}ed map is
non-Gaussian. In sidelobe contaminated region, it is rather
obvious. To be unbiased, the recovery test should be in the predicted
central image region. We must compromise since we need equal treatment
({\sc clean} them simultaneously) for both the central image and the
artificial sources. We chose the region close to the central image
region to minimise this position dependency effect. To avoid confusion
with false alarms, we should avoid inject artificial sources on noise
spikes. This is easy to check without artificial source injection, and
for the small sample test the false alarms are negligible. For example,
for Gaussian noise, if you inspect 20000 beams, you will expect one
$4\sigma$ noise spike.

$\diamond$ There is no strict constraint on the expected central image
flux density distribution unless it violates the flux ratios allowed
by the cusped lens model, i.e., the central image should be less
bright than the saddle point image B ($S_C<S_B$). Of course, we need
to be objective since we can always resort to more complicated models
if something odd happens.

$\diamond$ A null hypothesis. Since we found no prominent image in the
central image region and there is no way to tell it apart from noise
spikes even if it is present, we claim non-detection of the central
image in our observation. We wish to derive an upper limit on the source
strength.

$\diamond$ A suboptimal strategy. The optimal statistic could be
derived according to the {\em Neyman-Pearson criterion} if we know the
accurate mathematical model of the noise of {\sc clean}
\citep{allen.02.phrvd}. However, in the absence of any noise analysis of the
{\sc clean} procedure itself \citep{cornwell.99.asp}, we
conservatively choose the maximum statistic (noise peak) from the
central image region in the recovery test as our suboptimal statistic,
accompanied by our null hypothesis. This statistic is taken as the
counting threshold in our recovery test.

Since we found from our ``ring test'' that false dismissals happened
with artificial sources less than 200 $\mu$Jy, we injected 20
artificial sources of 180 $\mu$Jy with varying positions close to the
central image region and did the recovery test. Most of them are
recovered with flux losses of $\sim$ 20-30\%, but two of them are
recovered with maximum fluxes lower than the 42 $\mu$Jy noise peak we
obtained from the central image region. \\

\noindent
(i) Frequentist upper limit: \\

This is an ``objective'' approach which is mainly based on the counting
frequency in our test. In fact, the standard {\em Neyman construction}
is a frequentist method. This method has been extensively used in weak
detection experiments in particle physics and gravitational wave
physics \citep{barlow.03.sppp, cranmer.03.sppp, abbott.04.phrvd}.

Let $D$ be the detection statistic, and $D_0$ the suboptimal statistic
of a targeted search in our test. If we inject artificial sources
greater than or equal to $S_0$ in the recovery test and we get a
fraction $C$ of trials having residuals above $D_0$, then we have a
frequentist upper limit $S_0(C)$ on the strength of the targeted
source, with confidence level $C$. This frequentist confidence level
should be calculated from the integrals of the probability
distribution function (PDF) $p(D|S_0)$: 

\begin{equation} 
C(S_0)=P(D\geqslant D_0|S_0)={\displaystyle \int_{D_0}^\infty p(D|S_0){\rm d}D
\over \displaystyle \int_0^\infty p(D|S_0){\rm d}D}.  
\end{equation}

\begin{figure}
\centering
\includegraphics[width=8cm]{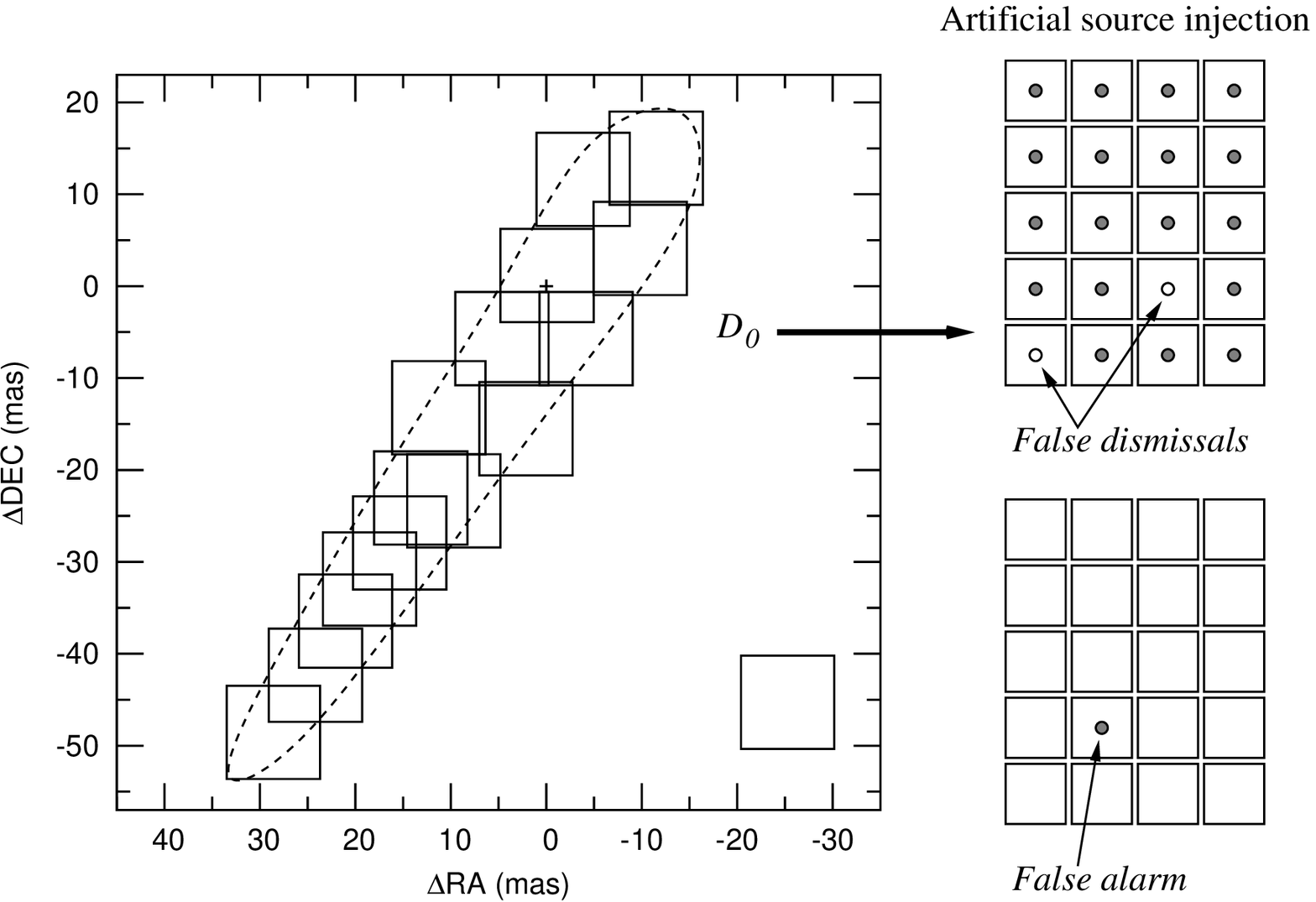}
\caption{Central image {\sc clean}ing and artificial source recovery
counting. The central image region is covered by $10\times10$~mas$^2$
{\sc clean} boxes. The statistics $D_0$ obtained from imaging is used
as the counting threshold. The right panel shows the possible false
dismissals with artificial source injection and false alarms without
artificial source injection in the real test.}
\label{fig:inject}
\end{figure}

In our test, we can count how many values of $D$ are greater than or
equal to $D_0$ and divide it by the total number of $D$ values. This
is illustrated in Fig.~\ref{fig:inject}. Our null and alternative
hypotheses are
\begin{align} 
H_0 &: D(S) \leqslant D_0, \notag\\
H_1 &: D(S) > D_0,
\end{align} 
and the false alarm and false dismissal probabilities are
\begin{align} 
\alpha(S) &= \int_{x \in H_1} p(x|H_0){\rm d}x, \notag\\
\beta(S)  &= \int_{x \in H_0} p(x|H_1){\rm d}x.
\end{align} 
We denote the detection probability $\gamma(S)=1-\alpha(S)-\beta(S)$, then the
confidence level will be the detection probability at a source
strength $S_0$ in this case
\begin{equation} 
C(S_0)=\gamma(S_0)=1-\alpha(S_0)-\beta(S_0)\simeq1-\beta(S_0).
\end{equation} 
Since we have 2 of 20 (remember the predicted central image region is
of about 20 beams) 180 $\mu$Jy artificial sources covered above 42
$\mu$Jy in our test, we will therefore say that if the source to be
detected is at 180 $\mu$Jy level, it should be detected in our map with a
flux over 42 $\mu$Jy, with a confidence level of $(20-2)/20 = 90\%$.

We have found from the recovery test that weak sources are more
likely to be {\sc clean}ed out, while strong sources more unlikely, i.e.,
\begin{align}
P(D \geqslant D_0|S<S_0) < P(D \geqslant D_0|S=S_0), \notag\\
P(D \geqslant D_0|S>S_0) > P(D \geqslant D_0|S=S_0),
\end{align}
then according to the {\em sum rule}
\begin{align}
P(D \geqslant D_0|S \leqslant S_0)< P(D \geqslant D_0|S=S_0), \notag\\
P(D \geqslant D_0|S \geqslant S_0)> P(D \geqslant D_0|S=S_0).
\end{align}
So if there is a true source of 180 $\mu$Jy of higher, the the
probability of getting an observation as small as we obtain or smaller
is only 10\% or less, i.e., to argue that there is a strong source
which happens to have a low fluctuation is very implausible. If we deny
the existence of a $\geqslant$180 $\mu$Jy source in this data, we will
be right at least 90\% of the time.

One should be aware that at different targeted source strength, we could
get different confidence levels. Moreover, our recovery test could be
biased because of the sampling size and position dependency. \\

\noindent
(ii) Bayesian (conditionalist) upper limit: \\

\begin{figure}
\centering
\includegraphics[width=8cm]{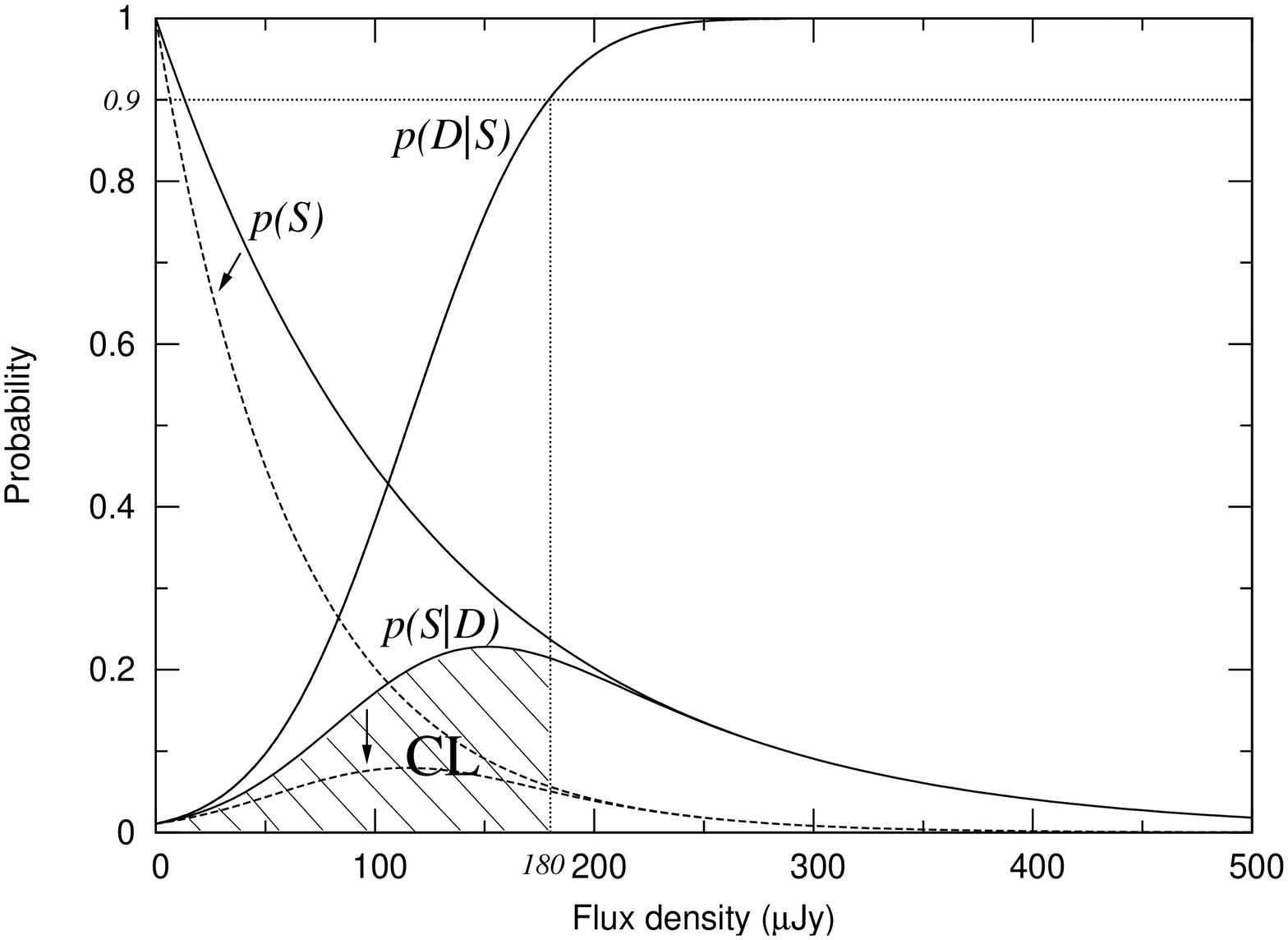}
\caption{Bayesian upper limit. The solid curves present the prior
  $p(S)$, likelihood $p(D|S)$ and the posterior probability $p(S|D)$;
  the dashed curves present another case of different prior and
  consequently different posterior probability of which the confidence
  level CL increases.}
\label{fig:bayes}
\end{figure}

The Bayesian approach has been consistently used in all statistical
studies, especially for hypothesis testing of theoretical predictions
provided we know the prior probability distributions of certain
physical quantities. This ``subjective'' approach (for informative
objective Bayesian interpretations, see \citealt{jaynes.03.bk}) which
always needs a prior knowledge of a certain probability distribution,
might not always be practical. The Bayesian confidence level
(``degree of belief'') of the upper limit is calculated from the
posterior probability $P(S \leqslant S_0|D \geqslant D_0)$. We can
invert the probability $P(D|S)$ according to {\em Bayes' theorem}
\begin{align}
P(S \leqslant S_0|D \geqslant D_0) & ={P(D \geqslant D_0|S \leqslant
  S_0)P(S \leqslant S_0) \over P(D \geqslant D_0)} \notag\\
 & = {\displaystyle \int_0^{S_0}p(S|D \geqslant D_0){\rm d}S \over
  \displaystyle \int_0^\infty p(S|D \geqslant D_0){\rm d}S} \notag\\
 & = {\displaystyle \int_0^{S_0}p(D \geqslant D_0|S)p(S){\rm d}S \over
  \displaystyle \int_0^\infty p(D \geqslant D_0|S)p(S){\rm d}S}.
\label{eq:bayes}
\end{align}
Now to calculate the posterior probability $P(S \leqslant S_0|D
\geqslant D_0)$, we need knowledge about the probability
distributions $P(S\leqslant S_0)$ and $P(D \geqslant D_0)$. However,
the probability distribution of real source strength $p(S)$ is not
known to us. Naturally we could assume a uniform (improper) prior here
$p(S)=constant$. Even though we do not have much information about the
data bias due to {\sc clean}ing, we can integrate the {\em total
probability} of $P(D)$ decomposed by the ``cause'' probability $p(S)$,
see Eq.~\ref{eq:bayes}. After normalisation, the probability $P(S
\leqslant S_0|D \geqslant D_0)$ calculation becomes the probability
$P(D \geqslant D_0|S \leqslant S_0)$ calculation, i.e., counting the
number of recoveries in the recovery test. Indeed, this uniformly most
powerful (UMP) test does exist if we assume a bound $S$ range, i.e.,
the integration $p(D|S)$ over $(0,\infty)$ converges (for more
description of tests in absence of a prior distribution, see
\citealt{selin.66.bk}). However, the Bayesian confidence level from
the UMP test will be very different from the frequentist one. Furthermore,
the truth is that the unknown prior source strength distribution we
assumed will affect the confidence level determination
significantly. Since we have explored from our recovery test that at
higher flux density level the recovery probability is higher, at lower
flux density level the recovery probability is lower, we assume a
monotonic continuous PDF of $p(D|S)$ which is illustrated in
Fig.~\ref{fig:bayes}. We can see from Fig.~\ref{fig:bayes} and
Eq.~\ref{eq:bayes} that the probability density $p(S|D)$ is totally
dependent on the profile of prior source PDF $p(S)$.  The unnormalised
inverse probability density $p(S|D)$ is the product of $p(D|S)$ and
$p(S)$. So the Bayesian confidence level of the upper limit will be
the fraction of the shaded part in the whole region below the curve,
i.e., the cumulative probability $P(S
\leqslant S_0|D)$. In the illustrated case, the confidence level is
definitely less than 90\%. Obviously, we can manipulate the shape of
$p(S)$ to make the confidence level arbitrary.

In brief, if we want to set up an ``objective'' upper limit from an
experiment, without assuming a Bayesian prior, we should subscribe to
the frequentist point of view. The approach taken in our paper is indeed a
frequentist approach.

\bibliographystyle{my}
\bibliography{refs}

\end{document}